\documentclass[aps,prd,preprintnumbers,superscriptaddress,nofootinbib,twocolumn]{revtex4-1}

\usepackage[dvipdfmx]{graphicx}
\usepackage{here}
\usepackage{braket}
\usepackage{feynmf}
\usepackage{xcolor}
\usepackage{bm,latexsym,amsmath,amssymb,amsfonts,mathrsfs}
\usepackage[colorlinks=true,urlcolor=teal,citecolor=orange,linkcolor=violet]{hyperref}
\usepackage{slashed}
\usepackage{framed}

\newcommand*{\ba}{\begin{eqnarray}}
\newcommand*{\ea}{\end{eqnarray}}
\newcommand*{\bb}{\begin{framed}}
\newcommand*{\eb}{\end{framed}}

\newcommand*{\mpl}{M_{\rm Pl}}
\newcommand{\simgt}{\lower.5ex\hbox{$\; \buildrel > \over \sim \;$}}
\newcommand{\simlt}{\lower.5ex\hbox{$\; \buildrel < \over \sim \;$}}

\newcommand*{\calA}{{\cal A}}
\newcommand*{\calB}{{\cal B}}
\newcommand*{\calC}{{\cal C}}
\newcommand*{\calD}{{\cal D}}

\newcommand*{\calF}{{\cal F}}
\newcommand*{\calG}{{\cal G}}

\newcommand*{\calL}{{\cal L}}

\newcommand{\nn}{\nonumber \\}

\begin{document}

\title{Exploring spin of ultralight dark matter with gravitational wave detectors}

\author{Yusuke \sc{Manita}}
\email{manita@tap.scphys.kyoto-u.ac.jp}
\affiliation{Department of Physics, Kyoto University, Kyoto 606-8502, Japan}

\author{Hiroki \sc{Takeda}}
\affiliation{Department of Physics, Kyoto University, Kyoto 606-8502, Japan}

\author{Katsuki \sc{Aoki}}
\affiliation{Center for Gravitational Physics and Quantum Information, Yukawa Institute for Theoretical Physics, Kyoto University, 606-8502, Kyoto, Japan}

\author{Tomohiro \sc{Fujita}}
\affiliation{Waseda Institute for Advanced Study, Shinjuku, Tokyo 169-8050, Japan}

\author{Shinji \sc{Mukohyama}}
\affiliation{Center for Gravitational Physics and Quantum Information, Yukawa Institute for Theoretical Physics, Kyoto University, 606-8502, Kyoto, Japan}
\affiliation{Kavli Institute for the Physics and Mathematics of the Universe (WPI), The University of Tokyo, 277-8583, Chiba, Japan}

\preprint{KUNS-2979, YITP-23-122, IPMU23-0034}

 \date{\today}

\begin{abstract}
We propose a novel method for distinguishing the spin of ultralight dark matter (ULDM) using interferometric gravitational wave detectors. ULDM can 
be a bosonic field of spin-0, 1, or 2, and each induces distinctive signatures in signals. We find that the finite-time traveling effect
causes a dominant signal for spin-0 and spin-1 ULDM,
while not for spin-2. 
By using 
overlap reduction functions (ORF) of multiple detectors, 
we can 
differentiate between the spins of ULDM. 
Furthermore, we point out that the current constraint on the coupling constant of spin-1 ULDM to baryons becomes 30 times weaker when the finite-time light-travel effect on the ORF is taken into account.
\end{abstract}

\maketitle


\section{Introduction}

Dark matter is an unknown non-baryonic component accounting for approximately 20\% of energy density in the current universe. Various dark matter models with different masses have been proposed so far. The ultralight dark matter (ULDM) is the lightest model with a mass range of about $10^{-22}-1~{\rm eV}$. While the ULDM must be composed of bosons because such a light fermion cannot form a high-density structure like dark matter halo due to the Pauli blocking~\cite{Tremaine:1979we}, a bosonic field with any integer spin can be a ULDM candidate in principle. 

The spin-0 (scalar) ULDM is motivated by QCD axion~\cite{Weinberg:1977ma,Wilczek:1977pj}, as well as superstring theory~\cite{Arvanitaki:2009fg}. Superstring theory expects various scalar fields such as axion, moduli, and dilaton, any of which can be a ULDM candidate~\cite{Arvanitaki:2014faa}. The spin-1 (vector) ULDM, which is also called dark photon, was originally introduced by Ref.~\cite{Holdom:1985ag} and string theory anticipates a light mass dark photon~\cite{Goodsell:2009xc}. In addition, a massive spin-2 field, which arises as a low-energy modification of Einstein's gravity, is equally a viable candidate for ULDM~\cite{Aoki:2017cnz, Marzola:2017lbt, Manita:2022tkl, Gorji:2023cmz}. 

For each spin and mass, various methods to search for the ULDM have been suggested. The mass range of about $10^{-14}-10^{-10}~{\rm eV}$ can be explored by the ground-based interferometric gravitational wave detectors, which is the main focus of this paper. The spin-0 ULDM generates detectable signals when it changes the fundamental constants such as the fine structure constant or the proton-electron mass ratio~\cite{Arvanitaki:2014faa, Stadnik:2014tta, Morisaki:2018htj}. The spin-1 ULDM coupled to the $U(1)_B$ or $U(1)_{B-L}$ current leaves signals by displacing the mirror in the detectors~\cite{Pierce:2018xmy}. In the case of the spin-2 ULDM, it couples to the energy-momentum tensor of standard model particles due to its gravitational origin and gives signals in the same way as the gravitational waves~\cite{Armaleo:2020efr,Manita:2022tkl}. 

The spin-0 and spin-1 ULDM have been explored by using real data from gravitational wave detectors. For the spin-0 ULDM, constraints on the coupling constants with the standard model particles have been provided by GEO600~\cite{Dooley:2015fpa} and LIGO's O3 run~\cite{Fukusumi:2023kqd}. On the other hand, constraints on the spin-1 ULDM have been obtained by analyzing the data in LIGO's O1 run~\cite{Guo:2019ker} and O3 run~\cite{LIGOScientific:2021ffg}.

While the spin-2 ULDM causes differential motions of the mirrors in detectors as with gravitational waves, the spin-1 and spin-0 ULDM induce common motions. As a result, the signal from the spatial displacement of the mirrors is well suppressed in the spin-0 and spin-1 ULDM, and the ``finite-time traveling effect" becomes dominant if the ULDM mass is $m\gtrsim v/L~(\simeq 50{\rm Hz}$ for LIGO) where $v$ and $L$ denote the velocity of ULDM and the arm length of detector~\cite{Morisaki:2018htj,Morisaki:2020gui}.

We propose a method to distinguish ULDM signals of spin-0, spin-1, and spin-2 signals by using cross-correlation analysis with multiple detectors. The difference in mirror motions is imprinted on the overlap reduction function (ORF), which is determined by the relative position and orientation of the two detectors. Notably, we find that the ORF for the spin-0 and 1 ULDM signals have a quite different dependency on the relative position and orientation compared to that for spin-2, owing to the finite-time traveling effect in the spin-0 and 1 ULDM signals. Note that the differences in power spectra in ULDM can be also used to distinguish spins~\cite{Miller:2022wxu}. 

Furthermore, our results suggest that the current upper bound on the coupling constant $\epsilon_B$ for the spin-1 ULDM should be weakened because the analysis in~\cite{LIGOScientific:2021ffg} uses the ORF designed for gravitational waves, not for the spin-1 ULDM with the finite-time traveling effect. Specifically, the constraint on $\epsilon_B^2$ from LIGO's and Virgo's O3 run~\cite{LIGOScientific:2021ffg}, becomes approximately 30 times less stringent.

This paper is organized as follows. In Sec.~\ref{sec:uldmsignal}, we review the ULDM models and their detectable signals in gravitational wave detectors. In Sec.~\ref{sec:cross-correlation}, we develop the method for the cross-correlation analysis of spin-1 and spin-2 ULDM signals. In Sec.~\ref{sec:ORF}, we show that the finite-time traveling effect induces an ORF whose position and orientation dependence on detectors are different from those for spin-2. We also discuss the current and future constraints on spin-0, spin-1 and spin-2 ULDM. In Sec.~\ref{sec:ExploringSpinofULDM}, we examine the distinguishability of the signals originating from spin-0, spin-1, and spin-2 ULDM based on the difference of their ORFs. Finally, Sec.~\ref{sec:summary} is devoted to a summary and discussion. Throughout this paper, we use the unit of $c=\hbar=1$ and the notation $M_{\rm Pl}=1/\sqrt{8\pi G}$ where $G$ is Newton's constant.

\section{ULDM search in gravitational wave detectors}
\label{sec:uldmsignal}

In this section, we review the models of ULDM and explain how they can be probed by gravitational wave detectors. 

\subsection{Spin-2 ULDM Search}

Spin-2 dark matter is a class of dark matter models that arises from bigravity~\cite{Aoki:2016zgp,Babichev:2016bxi,Babichev:2016hir} or multi-gravity theories~\cite{GonzalezAlbornoz:2017gbh}. In the ultralight mass range (lighter than a few eV scales), there are viable generation mechanisms such as the misalignment mechanism~\cite{Marzola:2017lbt}, production from primordial magnetic fields~\cite{Aoki:2017cnz}, or production from the anisotropic universe~\cite{Manita:2022tkl}, which can account for all or a part of dark matter component of the universe.

In the present Universe, such dark matter can be modeled as a superposition of tensor non-relativistic waves $\varphi_{\mu\nu}(k)$ where $k$ is the four-momentum satisfying $k^2=-m^2$. For a plane wave, the linearized equations of motion for the massive graviton around the vacuum yield the relation $k_\mu\varphi^{\mu}{}_{\nu}=0,~\varphi^\mu{}_\mu=0$.
Since we are interested in the non-relativistic waves $k^0 \simeq m,~ k^i \simeq 0$, we have $\varphi^0{}_\nu \simeq 0$ and $\varphi^i{}_i \simeq 0$. Hence, we only consider the traceless part of $\varphi_{ij}$. The spin-2 ULDM (dark graviton) can be modeled by 
\begin{align}
    \Phi_{ij}(t,\bm{x}) 
    &=\int d^3\bm{k}\,\varphi_{ij}(\bm{k}) 
    \nn
    &\approx \sum_{\lambda\in \Lambda_{\rm DG}}\int d^3\bm{k}\,\tilde{\varphi}^{\lambda}(\bm{k})
    e_{ij}^{\lambda}(\hat{\bm{\Omega}})
\cos(\omega t-\bm{k}\cdot \bm{x}+\theta)\,,
    \label{eq:superposition_spin2}
\end{align}
where each plane wave $\varphi_{ij}(\bm{k})$ have been defined by
\begin{align}
    \varphi_{ij}(\bm{k})=\sum_{\lambda\in \Lambda_{\rm DG}}\tilde{\varphi}^{\lambda}(\bm{k})
    e_{ij}^{\lambda}(\hat{\bm{\Omega}})\cos(\omega t-\bm{k}\cdot \bm{x}+\theta)\,.
\end{align}
Since spin-2 ULDM is composed of non-relativistic waves, $\omega$ and $\bm{k}$ can be approximated as 
\begin{align}
    \omega \simeq m\left(1+\frac{v^2}{2}\right)\,,\quad
\bm{k} \simeq m v\hat{\bm{\Omega}}\,,
\label{eq:omega_nonrela}
\end{align}
where $v, \hat{\bm{\Omega}},$ and $\lambda$ are the speed of dark matter, its direction, and the polarization mode, respectively. 
 $\Lambda_{\rm DG}$ represents the set of possible polarization modes of the spin-2 ULDM.
For each wave, 
the amplitude $\tilde{\varphi}^{\lambda}(\bm{k})$ follows an appropriate distribution, such as the Maxwell distribution. 
However, the phase $\theta$ is a random variable that adheres to a uniform distribution for each $\bm{k}$ and $\lambda$. 

Since they do not affect our result significantly, we will neglect these randomnesses and assume a constant speed as
\begin{align}
    \Phi_{ij}(t,\bm{x})&=\sum_{\lambda\in\Lambda_{\rm DG}} \int_{S^2} d^2\hat{\bm{\Omega}}\,\tilde{\varphi}^\lambda(\hat{\bm{\Omega}}) e^{\lambda}_{ij}(\hat{\bm{\Omega}}) \nn
    &\quad\times\cos\left(\omega t-mv\hat{\bm{\Omega}}\cdot\bm{x}+\theta\right)\,,
\end{align}
where $\theta$ denotes a constant phase shift. In practice, the velocities of the plane waves constituting ULDM can take on a variety of values according to a certain distribution such as standard halo model~\cite{Evans:2018bqy}. However, within the scope of our present study, the variance of these velocities is not important and is therefore neglected.

The maximum number of polarization modes is five, denoted by $\Lambda_{\rm DG}=\{+,\times,V_x,V_y,S\}$, and $e^\lambda_{ij}$ are the corresponding polarization bases. To define these polarization tensors, we choose an orthonormal coordinate system with the bases $\{\hat{\bm{m}},\hat{\bm{n}},\hat{\bm{\Omega}}\}$. Here, $\hat{\bm{\Omega}}$ is defined as $\hat{\bm{\Omega}}\equiv\bm{k}/|\bm{k}|$, and $\hat{\bm{m}}$ and $\hat{\bm{n}}$ are unit orthogonal vectors on the polarization plane defined by $\hat{\bm{\Omega}}$. Due to the traceless condition, the polarization bases are given by
\begin{align}
    e^{+}_{ij} &=\hat{m}_i \hat{m}_j-\hat{n}_i \hat{n}_j \,,
    \\
    e^{\times}_{ij} &=\hat{m}_i\hat{n}_j+\hat{n}_i\hat{m}_j \,,
    \\
    e^{V_x}_{ij} &=\hat{m}_i\hat{\Omega}_j+\hat{\Omega}_i\hat{m}_j \,,
    \\
    e^{V_y}_{ij} &=\hat{n}_i\hat{\Omega}_j+\hat{\Omega}_i \hat{n}_j \,,
    \\
    e^{S}_{ij} &=\frac{1}{\sqrt{3}}\left(\hat{m}_i\hat{m}_j+\hat{n}_i\hat{n}_j-2\hat{\Omega}_i\hat{\Omega}_j \right)\,.
    \label{eq:esij}
\end{align}

Note that the number of polarization modes depends on the underlying theoretical framework. For instance, the Hassan-Rosen bigravity theory encompasses all five polarization modes: two tensor modes $e^+_{ij},e^\times_{ij}$, two vector modes $e^{V_x}_{ij},e^{V_y}_{ij}$, and one scalar mode $e^S_{ij}$~\cite{Hassan:2011zd}. On the other hand, the Minimal Theory of Bigravity includes only the two tensor modes $e^+_{ij}$ and $e^\times_{ij}~$\cite{DeFelice:2020ecp}.

The spin-2 dark matter universally couples to the matter fields through
\begin{align}
    \calL_{\rm int}=\frac{\alpha}{\mpl}\Phi_{ij}T_{\rm m}^{ij}\,,
\end{align}
where $\alpha$ is a dimensionless constant and $T_{\rm m}^{ij}$ is the spatial component of the energy-momentum tensor of matter fields. The coupling is exactly the same as the coupling to gravitational waves. In other words, when we ignore all gravitational effects other than the spin-2 dark matter, the matter fields propagate on the metric $g_{\mu\nu}=\eta_{\mu\nu}+ \alpha \Phi_{\mu\nu}/\mpl$ with $\Phi_{0\mu} \simeq 0$. The oscillation of $\Phi_{ij}$ then causes a motion of mirrors as with the standard gravitational waves.

In order to investigate the response of the detector, we adopt the proper reference frame to scrutinize the motion of the test mass mirrors. With the center of mass of two mirrors set to $x=0$, the motion of the test mass mirror at $x^i$ is described by the geodesic deviation equation
\begin{align}
    \frac{d^2 x^j}{dt^2}=-R_{j0k0}x^k\,,
    \label{eq:geodesic}
\end{align}
where $R_{j0k0}$ represents the components of the Riemann curvature of the metric $g_{\mu\nu}=\eta_{\mu\nu}+ \alpha \Phi_{\mu\nu}/\mpl$. Note that the proper reference frame is viable only when $\omega L \ll 1$ with the arm length $L$. This condition is generally satisfied for ground-based detectors such as LIGO. In the subsequent discussions, we focus only on ground-based detectors and assume $\omega L \ll 1$. 

Thanks to the transverse condition $\Phi_{0\mu}=0$, the $j0k0$ components of the linearized Riemann curvature are expressed by
\begin{align}
    R_{j0k0}\simeq-\frac{\alpha}{2\mpl}\ddot{\Phi}_{jk}\,.
\end{align}
We consider an arm directed along $\hat{\bm{X}}$, where $x^j=x\hat{X}^j$. For simplicity, the two test mass mirrors are assumed to have identical masses, thus the origin of the coordinate is the midpoint of the arm in the absence of spin-2 ULDM. Consequently, the geodesic equation \eqref{eq:geodesic} becomes
\begin{align}
    \frac{d^2x}{dt^2}=-\frac{\alpha x}{2\mpl}\ddot{\Phi}_{jk}\hat{X}^{j}\hat{X}^{k}\,.
    \label{eq:geodesic_deviation_x}
\end{align}
We denote the coordinates of the input and end mirror as 
\begin{align}
    x_{\rm i}(t)=-\frac{L}{2}+\delta x_-(t)\,,
    \quad
    x_{\rm e}(t)=\frac{L}{2}+\delta x_+(t)\,.
    \label{eq:coordinate_mirror}
\end{align}
Then, \eqref{eq:geodesic_deviation_x} can be perturbatively solved as
\begin{align}
    \delta x_\pm&=\mp\frac{\alpha L}{4\mpl}\Phi_{jk}\hat{X}^j \hat{X}^k\,.
\end{align}
Thus, the input and end mirrors move in different directions. In other words, spin-2 ULDM generates a differential motion in two mirrors as illustrated in Fig.~\ref{fig:illustration_mirror}.

In the gravitational wave detector, these displacements of the mirrors are converted into a phase shift of a laser. In the case of an L-shaped detector, the signal is converted into a phase difference as
\begin{align}
    h(t) = \frac{1}{2\pi\nu}\frac{\phi(t;\hat{\bm{X}})-\phi(t;\hat{\bm{Y}})}{2 L}\,,
\label{eq:h_DP}
\end{align}
where $\nu$ is the laser frequency, $\hat{\bm{X}}$ and $\hat{\bm{Y}}$ are the unit vectors along each arm, and $\phi(t;\hat{\bm{X}})$ is the phase shift of the laser beam during the round trip of the $\hat{\bm{X}}$-arm. 

The round-trip phase shift is given by
\begin{align}
    \phi(t;\hat{\bm{X}})=2\pi\nu(t-T_r)+\phi_0\,,
\end{align}
where $\phi_0$ is a constant phase, and $T_r$ is the round-trip time, expressed by
\begin{align}
    T_r=-x_{\rm i}(t)+2x_{\rm e}(t-L)-x_{\rm i}(t-2L)\,.
\end{align}
Therefore, the phase shift of the laser, under the existence of the spin-2 ULDM, is
\begin{align}
     \phi(t;\hat{\bm{X}})=-2\pi \nu(\delta L_{\rm time}+\delta L_{\rm space})+2\pi \nu(t-2L)+\phi_0\,,
     \label{eq:phitu}
\end{align}
where 
\begin{align}
    \delta L_{\rm time} &= -\delta x_-(t)+2\delta x_-(t-L)-\delta x_-(t-2L)
    \nn
    &= \frac{\alpha L }{\mpl}\sin^2\left(\frac{\omega L}{2}\right)\hat{X}^i \hat{X}^j\Phi_{ij}(t-L)\,,
    \label{eq:deltaLtime_s2}
    \\
    \delta L_{\rm space} &= 2(\delta x_+(t-L)-\delta x_-(t-L))
    \nn
    &= -\frac{\alpha L }{\mpl}\hat{X}^i \hat{X}^j\Phi_{ij}(t-L)\,.
    \label{eq:deltaLspace_s2}
\end{align}
The first effect is caused by the displacement of mirrors during the round-trip of the laser, which is known as the finite-time traveling effect. The second effect arises from the difference in displacement between the input and end mirrors, due to the differential motion of the mirror.

The ratio of these displacements is estimated as
\begin{align}
    \frac{\delta L_{\rm time}}{\delta L_{\rm space}}&\simeq \sin^2\left(\frac{mL}{2}\right)
    \nn
    &\approx 1.8 \times 10^{-5} \left(\frac{m}{2\pi\times 100{\rm Hz}}\right)^2 \left(\frac{L}{4{\rm km}}\right)^2
    \,.
    \label{Lratio2}
\end{align}
Thus, $\delta L_{\rm time}$ is negligible in the ground-based detectors.

Therefore, we obtain the expression of the spin-2 dark matter signal in the detector as
\begin{align}
    h(t) &= -\frac{\alpha}{\mpl} D^{ij}\Phi_{ij}(t-L)
    \nonumber \\
    &= \frac{\alpha}{\mpl}\sum_{\lambda\in\Lambda_{\rm DG}} \int_{S^2} d^2\hat{\bm{\Omega}}\tilde{\varphi}^\lambda(\hat{\bm{\Omega}}) F^{\lambda}(\hat{\bm{\Omega}}) 
    \nn
    &\quad\times\cos\left(\omega (t-L)+\theta\right)\,,
    \label{eq:spin2_signal}
\end{align}
where we have defined the antenna pattern function as
\begin{align}
    F^\lambda(\hat{\bm{\Omega}}) \equiv D^{ij}e^\lambda_{ij}(\hat{\bm{\Omega}})\,,
\end{align}
with the detector tensor 
\begin{align}
    D^{ij} \equiv \frac{\hat{X}^i \hat{X}^j-\hat{Y}^i \hat{Y}^j}{2}\,.
    \label{eq:detectortensor}
\end{align}
As shown in \eqref{eq:spin2_signal}, the signal for spin-2 ULDM is obtained in the same way as the signal for regular massless gravitational waves.

\subsection{Spin-1 ULDM Search}

\begin{figure*}
    \centering
    \includegraphics[width=\linewidth]{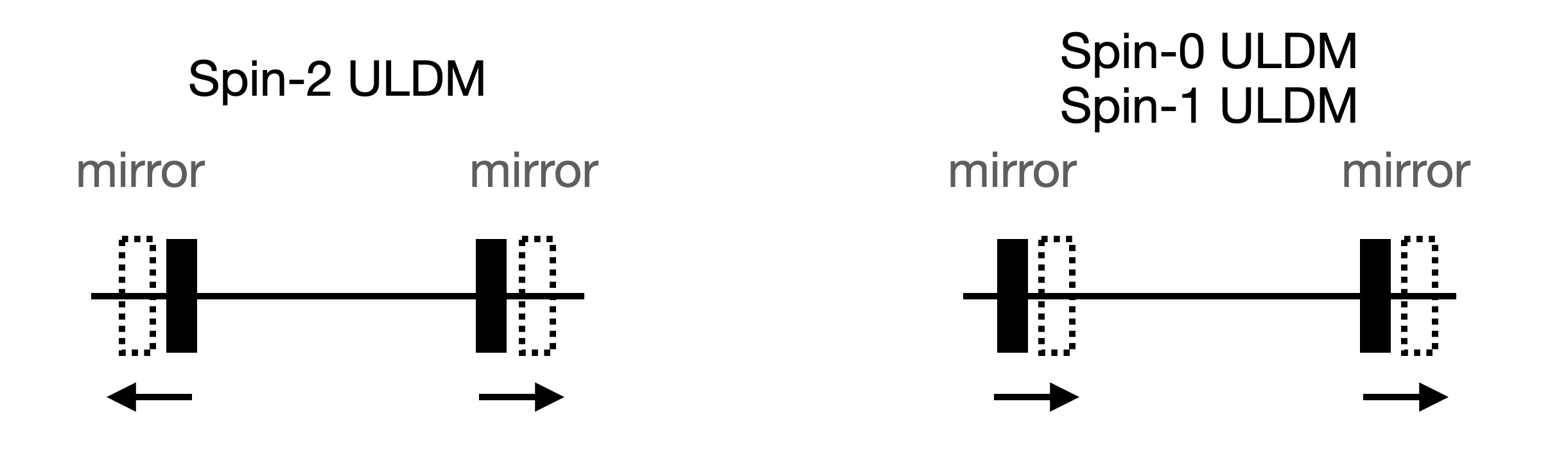}
    \caption{The schematic illustration of mirrors in a gravitational wave detector.
    The arrows show the motion of mirrors responding to spin-2 dark matter (left) and spin-0,1 dark matter (right). The former induces differential motion and the latter induces common motion of the mirrors.}
    \label{fig:illustration_mirror}
\end{figure*}

We then consider the spin-1 ULDM (dark photon)~\cite{Holdom:1985ag}. Several production mechanisms have been proposed for the non-thermal production of spin-1 ULDM in the early universe, for example, production from cosmic string~\cite{Long:2019lwl}, gravitational production during inflation~\cite{Graham:2015rva}, and so on.

The spin-1 ULDM is also constituted by a superposition of classical waves $A_{\mu}$. For a free massive spin-1 field, the equations of motion yield $k_{\mu}A^{\mu}=0$ and $k^2=-m^2$ so we can ignore the temporal component of $A_{\mu}$ in the non-relativistic limit. For simplicity, we employ a model with fixed $v$ and $\theta$ as
\begin{align}
    A_i(t,\bm{x}) &= \sum_{\lambda\in\Lambda_{\rm DP}}\int_{S^2} d^2\hat{\bm{\Omega}} A^\lambda(\hat{\bm{\Omega}}) e^\lambda_i(\hat{\bm{\Omega}})
    \nn
    &\quad\times\cos\left(\omega t-mv\hat{\bm{\Omega}}\cdot\bm{x}+\theta\right)\,.
    \label{eq:darkphotonfield}
\end{align}
We have defined the set of polarization as $\Lambda_{\rm DP}=\{x,y,z\}$, and corresponding polarization bases as
\begin{align}
    e^x_i = \hat{m}_i\,,
    \quad
    e^y_i = \hat{n}_i\,,
    \quad
    e^z_i = \hat{\Omega}_i\,.
\end{align}

Suppose that the spin-1 ULDM is a gauge boson associated with a symmetry of the standard model, $B$ or $B-L$. Then, the interaction between the spin-1 ULDM and the standard model particles is given by
\begin{align}
    \calL_{\rm int}=\epsilon_D e A_\mu j_D^\mu\,,
\end{align}
where $j_D^\mu$ with $D=B$ or $D=B-L$ represents the corresponding current of standard model particles and $\epsilon_D$ denotes the coupling constant, normalized by the elementary charge $e$. Consequently, in the presence of the spin-1 ULDM, test mass mirrors in the arm are subject to Coulomb-type force
\begin{align}
    \bm{F}&=-\epsilon_D e Q_D \dot{\bm{A}}(t,\bm{x})\,,
\end{align}
where $Q_D$ denotes the $U(1)_D$ charge of the mirror.

Projecting this force along the direction of the $\hat{\bm{X}}$-arm, the equation of motion for the test mass mirrors is given by
\begin{align}
    \frac{d^2x}{dt^2}=-\epsilon_D e \frac{Q_D}{M} \bm{\hat{X}}\cdot \dot{\bm{A}}\,,
\label{eq:eom_dp}    
\end{align}
where $M$ represents the mass of the mirror. Employing the same coordinate system as \eqref{eq:coordinate_mirror}, this equation of motion can be solved as
\begin{align}
    \delta x_{\pm}(t,\bm{x}) &= -\frac{\epsilon_D e}{\omega}\frac{Q_D}{M}\sum_{\lambda\in\Lambda_{\rm DP}}\int_{S^2}d^2\hat{\bm{\Omega}}\,A^\lambda(\hat{\bm{\Omega}})\hat{\bm{X}}\cdot\bm{e}^\lambda(\hat{\bm{\Omega}})
    \nn
    &\times\sin\left(\omega t-mv\hat{\bm{\Omega}}\cdot\bm{x}+\theta\right)\,.
\end{align}
Unlike the spin-2 ULDM, the spin-1 ULDM fluctuates the input and output mirrors in the same directions, as illustrated in Fig.~\ref{fig:illustration_mirror}. While the motion of mirrors is not a differential motion, the phase shift of the laser is still caused by the following two reasons. Firstly, as explained in the case of the spin-2 ULDM, the positions of the detectors at the emission are different from those at the arrival time due to the finite light-travel time, which causes a change in traveling distance of the laser even for the common motion of the mirrors~\cite{Morisaki:2018htj}. Secondly, when the distribution of the spin-1 ULDM has a spatial dependence, i.e.,~$v\neq 0$, the forces acting on detectors at different positions are different, leading to a displacement of the arm length.

The net displacement of the arm length is given by a sum of these two effects:
\begin{align}
    \delta L_{\rm time} &= -\frac{4\epsilon_D e}{\omega}\frac{Q_D}{M}\sin^2\left(\frac{\omega L}{2}\right)
    \nn
    &\quad\times\sum_{\lambda\in\Lambda_{\rm DP}}\int_{S^2}d^2\hat{\bm{\Omega}} A^\lambda(\hat{\bm{\Omega}})\hat{\bm{X}}\cdot\bm{e}^\lambda(\hat{\bm{\Omega}})
    \nn
    &\quad\times\sin\left(\omega (t-L)+\theta\right)\,,
    \label{eq:deltaLtime}
    \\
    \delta L_{\rm space} &= 2\epsilon_D evL\frac{Q_D}{M}
    \nn
    &\quad\times\sum_{\lambda\in\Lambda_{\rm DP}}\int_{S^2}d^2\hat{\bm{\Omega}}\,A^\lambda(\hat{\bm{\Omega}})(\hat{\bm{X}}\cdot\bm{e}^\lambda(\hat{\bm{\Omega}}))(\hat{\bm{X}}\cdot{\hat{\bm{\Omega}}})
    \nn
    &\quad\times\cos\left(\omega (t-L)+\theta\right)\,.
    \label{eq:deltaLspace}
\end{align}
The ratio of these displacements is roughly estimated as
\begin{align}
    &\frac{\delta L_{\rm time}}{\delta L_{\rm space}}\sim\frac{mL}{v}
    \nn
    &\approx 12.6\left(\frac{m}{2 \pi\times 100{\rm Hz}}\right)\left(\frac{L}{4 {\rm km}}\right) \left(\frac{200{\rm km/s}}{v}\right)\,.
\end{align}
Hence, the finite-time traveling effect $\delta L_{\rm time}$ dominates for the larger mass than $v/L~(\simeq 50$ Hz for LIGO)~\cite{Morisaki:2018htj}. This should be contrasted with the spin-2 ULDM \eqref{Lratio2}. In the case of spin-2, the leading effect of the spatial displacement $\delta L_{\rm space}$ is simply caused by the differential motion of mirrors and thus the finite-time traveling effect $\delta L_{\rm time}$ is subdominant. On the other hand, the spatial displacement $\delta L_{\rm space}$ appears as a result of a spatial inhomogeneity so $\delta L_{\rm space}$ is suppressed by a power of $v$. Hence, $\delta L_{\rm time}$ can give a dominant effect in the spin-1 ULDM.

For $\delta L_{\rm space}$, the angular dependence is the same as those of the scalar and vector polarization modes in the spin-2 ULDM. Therefore, using the detector tensor defined by \eqref{eq:detectortensor}, the corresponding antenna pattern function is expressed as
\begin{align}
    F^{\lambda}(\hat{\bm{\Omega}}) = 2 D^{ij}\Omega_i e^\lambda_j(\hat{\bm{\Omega}}),
    \quad
    (\lambda\in\Lambda_{\rm DP})\,.
\end{align}
In contrast, $\delta L_{\rm time}$ exhibits a different angular dependence. Introducing the detector vector as
\begin{align}
    D^i \equiv \hat{X}^i-\hat{Y}^i\,,
    \label{eq:detectorvector}
\end{align}
the corresponding antenna pattern functions are given by
\begin{align}
    G^\lambda(\hat{\bm{\Omega}})\equiv D^{i}e^\lambda_i(\hat{\bm{\Omega}})\,,
    \quad
    (\lambda\in\Lambda_{\rm DP})\,.
\end{align}

The phase shift of the laser also takes the form of \eqref{eq:phitu}, and the signal is expressed by \eqref{eq:h_DP}. Therefore, the spin-1 ULDM signal in gravitational waves detectors is represented by
\begin{align}
    h(t)=h_{\rm time}(t)+h_{\rm space}(t),
    \label{eq:signal_dark_photon_h}
\end{align}
where
\begin{align}
    h_{\rm time}(t)&=\frac{2\epsilon_D e}{\omega L}\frac{Q_D}{M}\sin^2\left(\frac{\omega L}{2}\right)
    \nn
    &\quad\times\sum_{\lambda\in\Lambda_{\rm DP}}\int_{S^2}d^2\hat{\bm{\Omega}}\,A^\lambda(\hat{\bm{\Omega}})G^\lambda(\hat{\bm{\Omega}})
    \nn
    &\quad\times\sin\left(\omega (t-L)+\theta\right)\,,
    \label{eq:htime}
    \\
    h_{\rm space}(t)&=-\epsilon_D ev\frac{Q_D}{M}
    \nn
    &\quad\times\sum_{\lambda\in\Lambda_{\rm DP}}\int_{S^2}d^2\hat{\bm{\Omega}}\,A^\lambda(\hat{\bm{\Omega}})F^\lambda(\hat{\bm{\Omega}})
    \nn
    &\quad\times\cos\left(\omega (t-L)+\theta\right)\,.
    \label{eq:hspace}
\end{align}

\subsection{Spin-0 ULDM Search}

The spin-0 ULDM (scalar dark matter) is motivated by solving the strong CP problem \cite{Peccei:1977hh} or superstring theory~\cite{Arvanitaki:2009fg}. They can be produced through the non-thermal mechanism, such as the misalignment mechanism~\cite{Abbott:1982af,Preskill:1982cy,Dine:1982ah}.   

As with the previous cases, the spin-0 ULDM is modeled by a superposition of a huge number of classical waves:
\begin{align}
    \phi(t, \bm{x}) &= \int_{S^2} d^2\hat{\bm{\Omega}} \varphi(\hat{\bm{\Omega}})
    \cos\left(\omega t-mv\hat{\bm{\Omega}}\cdot\bm{x}+\theta\right)\,.
    \label{eq:darkphotonfield}
\end{align}
We consider the effective interaction Lagrangian in the form of~\cite{Damour:2010rp,Arvanitaki:2014faa}
\begin{align}
    \calL_{\rm int}&=-\kappa\phi\left[\frac{d_e}{4e^2}F_{\mu\nu}F^{\mu\nu}-\frac{d_g\beta_3}{2g_3}G^A_{\mu\nu}G^{A\mu\nu}\right.
    \nn
    &\quad\left.-d_{m_e}m_e\bar{e}e-\sum_{i=u,d}(d_{m_i}+\gamma_{m_i}d_g)m_i\bar{\psi}_i\psi_i\right]\,,
\end{align}
where $F_{\mu\nu}$ is a field strength of the electromagnetic fields, and $G^A_{\mu\nu}$ is a gluon. $d_i$ are dimensionless parameters whereas $\beta_3$ and $\gamma_{m_i}$ are the QCD beta function and the anomalous dimensions of the up and down quarks, and $\kappa=\sqrt{4\pi G}$.

The motion of the mirrors in the arms of the detectors is described by the equation of motion~\cite{Morisaki:2018htj}
\begin{align}
    \frac{d^2 \bm{x}}{d t^2} \simeq-\kappa \alpha_A \bm{\nabla} \phi\,,
\label{eq:eomspin0}
\end{align}
where~\cite{Damour:2010rp}
\begin{align}
    \alpha_A \approx d_g^*+\left[-\frac{0.036}{A^{1/3}}-0.02\frac{(A-2Z)^2}{A^2}\right](d_{\hat{m}}-d_g)\,.
\end{align}
Here, $A$ and $Z$ are the mass number and the atomic number, and we have defined
\begin{align}
    d_g^*\approx d_g+0.093\left(d_{\hat{m}}-d_g\right)\,,
\end{align}
and
\begin{align}
    d_{\hat{m}} \equiv \frac{d_{m_u} m_u+d_{m_d} m_d}{m_u+m_d}\,.
\end{align}
Since the mirrors are largely composed of $\text{SiO}_2$, $\alpha_A$ can be approximately set as~\cite{Fukusumi:2023kqd}
\begin{align}
    \alpha_A\approx d_g+0.083(d_{\hat{m}}-d_g)\,.
\end{align}
In the following, we assume that all mirrors are composed by the same components for simplicity.

With the equation of motion \eqref{eq:eomspin0}, through a similar procedure as the spin-1 ULDM, the spin-0 ULDM gives the signal in the detectors as
\begin{align}
    h(t)=h_{\rm time}(t)+h_{\rm space}(t)\,,
    \label{eq:signal_dark_photon}
\end{align}
where
\begin{align}
    h_{\rm time}(t)&=-\frac{2\kappa \alpha_A v}{\omega L}\sin^2\left(\frac{\omega L}{2}\right)\int_{S^2}d^2\hat{\bm{\Omega}}\,\varphi(\hat{\bm{\Omega}})G^z(\hat{\bm{\Omega}})
    \nn
    &\quad\times\sin\left(\omega (t-L)+\theta\right)\,,
    \label{eq:htime_0}
    \\
    h_{\rm space}(t)&=-\kappa \alpha_A v^2\int_{S^2}d^2\hat{\bm{\Omega}}\,\varphi(\hat{\bm{\Omega}})F^z(\hat{\bm{\Omega}})
    \nn
    &\times\cos\left(\omega (t-L)+\theta\right)\,.
    \label{eq:hspace_0}
\end{align}
Thus, as with the spin-1 ULDM signal, the finite-time traveling effect dominates if the ULDM mass is larger than $v/L~(\simeq$ 50 Hz for LIGO). We also note that the antenna pattern functions are the same as those of the scalar polarization mode of the spin-1 ULDM.

\subsection{Distinguish ULDM Spin with Single Detector}

As we have discussed, the ULDMs of spin-0, 1, and 2 may all give detectable signals in gravitational wave detectors. We now turn to the question of whether we can distinguish the spin of ULDM if signals are detected. At first glance, it apears to be easy to distinguish because the signals are composed of a different set of polarization modes depending on the spin of ULDM. In practice, however, it is difficult to discriminate the spins by {\it a single detector}.  This is because, as explicitly shown in \eqref{eq:spin2_signal}, \eqref{eq:htime}, \eqref{eq:hspace}, \eqref{eq:htime_0}, and \eqref{eq:hspace_0}, the ULDM signal for any spin can be regarded as a plane wave oscillating at angular frequency $\omega$, and therefore their waveforms are completely identical.

It is possible to find a spin-dependence of signals when considering the distribution of $v$~\cite{Nakatsuka:2022gaf}. The signal in Fourier space is no longer monochromatic and has a width of about $\Delta f/f_0 \sim v^2 \sim 10^{-6}$ due to the c distribution. Since the shape of the power spectrum depends on the spin, the spin may be distinguishable if we can precisely determine the power spectrum by experiments. In fact, Ref.~\cite{Miller:2022wxu} has demonstrated that spin-0 and spin-1 ULDMs can be distinguished in simulations. However, in the mass region where the finite-time traveling effect is dominant, the shapes of the power spectrum for spin-1 and spin-2 are almost the same because both power spectrum has the identical power of $v$. This makes them challenging to be distinguished.

\section{Cross-correlation analysis}
\label{sec:cross-correlation}

Let us shift our focus to cross-correlation analysis using {\it multiple detectors}. Cross-correlation is a method widely utilized in the data analysis of gravitational waves, and it has been employed in searches for dark photons and scalar dark matter~\cite{Guo:2019ker, LIGOScientific:2021ffg, Morisaki:2020gui}. In cross-correlation searches, the dependence of the antenna pattern functions is integrated into the overlap reduction function (ORF). 
In this section, we will express the signal-to-noise ratio of the cross-correlation signal in terms of ORF and the parameters of ULDM.

\subsection{Optimal Filtering and Signal-to-Noise Ratio for Cross-correlation ULDM Search}

First, we briefly review optimal filtering in cross-correlation analysis and discuss the application to the ULDM search. The technique employed here is adapted from the stochastic gravitational wave background search outlined by \cite{Allen:1997ad}. 

We initially define the cross-correlation signal as
\begin{align}
    S_{IJ}\equiv\int_{-T/2}^{T/2}dt\int_{-T/2}^{T/2}dt's_I(t)s_J(t')Q(t-t')\,,
    \label{eq:cross-correlationalsignal}
\end{align}
where $s_I(t)$ is the output from a given detector with $I$ and $J$ representing distinct detectors. The symbol $T$ indicates the observation time, and $Q(t-t')$ is a filter function that will be optimized to maximize the value of $S_{IJ}$.

By the Fourier transformation, $s_I(t)$ and $Q(t)$ can be expanded as
\begin{align}
   s_I(t)&=\int_{-\infty}^{\infty}df \tilde{s}_I(f) e^{2\pi i f t}\,,
    \\
   Q(t-t')&=\int_{-\infty}^{\infty}df \tilde{Q}(f) e^{2\pi i f (t-t')}\,.
\end{align}
Note that, as $s_I(t)$ and $Q(t-t')$ are real numbers, their Fourier coefficients satisfy $\tilde{s}_I(f)^*=\tilde{s}_I(-f)$ and $\tilde{Q}(f)^*=\tilde{Q}(-f)$. Substituting these expressions, the cross-correlation signal $S$ can be reformulated as
\begin{align}
    S=\int_{-\infty}^{\infty} df \int_{-\infty}^{\infty} df'  \delta_T(f-f')\tilde{s}^*_I(f)\tilde{s}_J(f')\tilde{Q}(f'),
    \label{eq:S_fourier}
\end{align}
where we have defined $\delta_T(f)$ as
\begin{align}
    \delta_T(f)\equiv\int_{-T/2}^{T/2} dt~e^{2\pi i f t}=\frac{\sin(\pi f T)}{\pi f}\,.
    \label{eq:delta_T}
\end{align}
In the limit of $T\to\infty$, the function $\delta_T$ reduces to a Dirac's delta function, thus it can be regarded as a finite approximation of a Dirac's delta function. 
Note that the optimal filter function is contingent on the positions and orientations of the detectors and exhibits a rapid decline if the time difference $t-t'$ exceeds the spatial distance between the two detectors. Given that the typical observation time is significantly larger than the spatial distance between the detectors, the range of the time integral may be approximated by infinity:
\begin{align}
    \int^{\infty}_{-\infty}df\, \delta_T(f) F(f) \simeq F(0)
    \,, \label{deltaT}
\end{align}
for a function $F(f)$. We use \eqref{deltaT} throughout the calculations but keep the finiteness of $T$ for later convenience [see e.g.~\eqref{eq:mu_calc} below].

We define the Signal-to-Noise Ratio (SNR) of the cross-correlation signal as
\begin{align}
    {\rm SNR} \equiv \frac{\mu}{\sigma}\,,
    \label{eq:SNR}
\end{align}
where $\mu$ and $\sigma$ represent the mean and variance of $S$ respectively. 
We assume the detector output $s_I(t)$ can be decomposed as
\begin{align}
    s_I(t) = h_I(t) + n_I(t)\,.
\end{align}
where $h_I(t)$ denotes a ULDM signal and $n_I(t)$ denotes a detector noise, respectively.
Provided that noise is uncorrelated between different detectors, the correlation function of outputs reduces to $\left<\tilde{s}^{*}_I(f)\tilde{s}_J(f')\right>=\left<\tilde{h}^{*}_I(f)\tilde{h}_J(f')\right>$. For the ULDM signal, this can be assumed to be denoted with a certain function $F_{IJ}(f)$ by
\begin{align}
    \left<\tilde{h}^{*}_I(f)\tilde{h}_J(f')\right>=F_{IJ}(|f|)\delta_{T}(f-f')\,.
    \label{eq:hh}
\end{align}
Consequently, the mean value of the cross-correlation signal $\mu = \left<S\right>$ is evaluated as
\begin{align}
    \mu
    &=T\int_{-\infty}^{\infty}df F_{IJ}(|f|)\tilde{Q}(f)\,.
    \label{eq:mu_calc}
\end{align}
To get this equation, we have applied $\delta_T(0)=T$, as derived from \eqref{eq:delta_T}.

To estimate the variance of $S$, we assume that the magnitude of the detector noise is much larger than the amplitude of the ULDM signal. Then, the squared variance of $S$ can be evaluated as
\begin{align}
    \sigma^2 &\equiv\left<S^2\right>-\left<S\right>^2 \approx \left<S^2\right>
    \nonumber \\
    &\approx\int_{-\infty}^{\infty}df \int_{-\infty}^{\infty}df' \int_{-\infty}^{\infty}dk \int_{-\infty}^{\infty}dk'
    \nonumber \\
    &\quad\times\delta_T(f-f')\delta_T(k-k')
    \nonumber \\
    &\quad\times\left<\tilde{n}^*_I(f)\tilde{n}_{J}(f')\tilde{n}^*_{I}(k)\tilde{n}_J(k')\right>\tilde{Q}(f')\tilde{Q}(k').
    \label{eq:sigma2_calc}
\end{align}
We also assume that $n_I(f)$ is an uncorrelated Gaussian random field with the noise power spectrum $S_{n,I}(f)$, that is,
\begin{align}
    \left<\tilde{n}_{I}^{*}(f)\tilde{n}_{I}(f')\right>=\frac1 2 S_{n,I}(|f|)\delta(f-f').
    \label{eq:noise_spectrum}
\end{align}
This allows us to express the four-point function of the noise in the integral in \eqref{eq:sigma2_calc} in terms of the power spectrum as
\begin{align}
    &\left<\tilde{n}^*_I(f)\tilde{n}_I(f')\tilde{n}^*_J(k)\tilde{n}_J(k')\right>
    \nonumber \\
    &=\left<\tilde{n}^*_I(f)\tilde{n}_I(k)\right>\left<\tilde{n}^*_J(f')\tilde{n}_J(k')\right>
    \nonumber \\
    &=\frac{1}{4}S_{n,I}(|f|)S_{n,J}(|k|)\delta_{T}(f-f')\delta_{T}(k-k')\,,
    \label{eq:four_point_noise}
\end{align}
where we have assumed $I\neq J$.
As a result, we find
\begin{align}
    \sigma^2
    &=\frac{1}{4}\int_{-\infty}^{\infty}df\int_{-\infty}^{\infty}df'S_{n,I}(|f|)S_{n,J}(|f|)\delta_T(f-f')^2|\tilde{Q}(f')|^2
    \nonumber \\
    &\approx\frac{T}{4}\int_{-\infty}^{\infty}dfS_{n,I}(|f|)S_{n,J}(|f|)|\tilde{Q}(f)|^2,
    \label{eq:sigma2_eval}
\end{align}
where \eqref{deltaT} and $\delta_T(0)=T$ are used to get the second line.

Substituting \eqref{eq:mu_calc} and \eqref{eq:sigma2_eval} into the definition of the SNR \eqref{eq:SNR}, we obtain
\begin{align}
    {\rm SNR}=2\sqrt{T}\frac{\int_{-\infty}^{\infty}dfF_{IJ}(|f|)\tilde{Q}(f)}{\left[\int_{-\infty}^{\infty}dfS_{n,I}(|f|)S_{n,J}(|f|)|\tilde{Q}(f)|^2\right]^{1/2}}\,.
\end{align}
To identify the optimal filter that maximizes the SNR, we introduce the inner product $(\cdot,\cdot)$ in the functional space as
\begin{align}
    (A,B)\equiv\int_{-\infty}^{\infty} df A(f)^*B(f)S_{n,I}(|f|)S_{n,J}(|f|)\,,
    \label{eq:inner_product}
\end{align}
where $A$ and $B$ are arbitrary functions of $f$. With this inner product, the SNR can be rewritten as
\begin{align}
    {\rm SNR}=2\sqrt{T}\frac{\left(\tilde{Q}(f),\frac{F_{IJ}(|f|)}{S_{n,I}(|f|)S_{n,J}(|f|)}\right)}{\left(\tilde{Q}(f),\tilde{Q}(f)\right)^{1/2}}\,.
    \label{eq:SNR2}
\end{align}
This can be regarded as the projection of $\frac{F_{IJ}(|f|)}{S_{n,I}(|f|)S_{n,J}(|f|)}$ in the direction of $\tilde{Q}(f)$ in the functional space. Hence, the optimal filter function $\tilde{Q}(f)$, which maximizes the SNR, is
\begin{align}
    \tilde{Q}(f)=K\frac{F_{IJ}(|f|)}{S_{n,I}(|f|)S_{n,J}(|f|)}\,,
    \label{eq:Q_opt}
\end{align}
where $K$ is an arbitrary constant. Therefore, the SNR under the optimal filtering is given by
\begin{align}
    {\rm SNR}=2\sqrt{T}\left[\int_{-\infty}^{\infty}df\frac{|F_{IJ}(|f|)|^2}{S_{n,I}(|f|)S_{n,J}(|f|)}\right]^{1/2}\,.
    \label{eq:SNR_opt}
\end{align}
As we will see, the correlation for the ULDM signal can be approximated by
\begin{align}
    F_{IJ}(f)\approx\calF_{IJ}\delta\left(|f|-\frac{\omega}{2\pi}\right)\,,
\end{align}
with $\calF$ which is independent of $f$. Note that when the observational time exceeds the coherent time of ULDM, the stochastic effect cannot be negligible which can be taken into account by replacing $T$ with $T_{\rm eff}$ defined by~\cite{Budker:2013hfa,Nakatsuka:2022gaf}
\begin{align}
    T_{\rm eff}=
    \begin{cases}
        T & (T\lesssim\tau_{\rm coh})\,, \\
        \sqrt{\tau_{\rm coh}T} & (T\gtrsim\tau_{\rm coh})
        \,.
    \end{cases}
\end{align}
All in all, SNR for the cross-correlation search of ULDM is expressed in the form
\begin{align}
    {\rm SNR}\approx\frac{2\sqrt{2}\calF_{IJ} T_{\rm eff}}{\sqrt{S_{n,I}(\frac{m}{2\pi})S_{n,J}(\frac{m}{2\pi})}}\,.
    \label{eq:SNR_fomula}
\end{align}

\subsection{Cross-correlation Search for Spin-2 ULDM}
\label{sec:cross_correlation_spin2}

For the spin-2 ULDM signal \eqref{eq:spin2_signal}, the Fourier transformation yields
\begin{align}
    \tilde{h}(f)&=\frac{\alpha}{2\mpl} 
    \sum_{\lambda\in\Lambda_{\rm DG}} \int_{S^2} d^2 \hat{\bm{\Omega}} ~ \tilde{\varphi}^\lambda(\hat{\bm{\Omega}}) F^\lambda(\hat{\bm{\Omega}})
    \nn
    &+\left[\delta\left(f-\frac{m}{2\pi}\right)e^{i\omega L+i\theta}+\delta\left(f+\frac{m}{2\pi}\right)e^{-i\omega L-i\theta}\right]\,.
    \label{eq:htilde_f}
\end{align}
In the galaxy, the ULDM is expected to be approximately isotropic, non-polarized, and stationary. Consequently, the two-point function of $\varphi_\lambda(\hat{\bm{\Omega}})$ is given by
\begin{align}
    \left<\tilde{\varphi}_\lambda(\hat{\bm{\Omega}})\tilde{\varphi}_{\lambda'}(\hat{\bm{\Omega}}')\right>=H_{\rm DG}^\lambda \delta_{\lambda\lambda'}\delta^2\left(\hat{\bm{\Omega}},\hat{\bm{\Omega}}'\right)
    \,,\label{eq:2pt_spin2}
\end{align}
where $\delta^2\left(\hat{\bm{\Omega}},\hat{\bm{\Omega}}'\right)$ is the delta function on the celestial sphere and $H_{\rm DG}^\lambda$ are non-negative real constants. By using this correlation, the energy density of spin-2 ULDM is
\begin{align}
    \rho_{\rm DG}=\frac{1}{4}\left<\dot{\tilde{\varphi}}_{ij}(t,\bm{x})\dot{\tilde{\varphi}}^{ij}(t,\bm{x})\right>=\pi \omega^2\sum_{\lambda\in\Lambda_{\rm DG}} H_{\rm DG}^\lambda\,.
    \label{eq:rhodg}
\end{align}
This indicates that the energy density of each polarization component is given by $\pi \omega^2 H_{\rm DG}^\lambda$. Therefore, we define the ratio of each polarization component as
\begin{align}
    \Omega^\lambda_{\rm DG}\equiv\frac{\pi \omega^2 H_{\rm DG}^\lambda}{\rho_{\rm DG}}\,,
    \label{eq:omega_dg}
\end{align}
for $\lambda\in\Lambda_{\rm DG}$. According to \eqref{eq:rhodg}, the sum of these fractions equals one:
\begin{align}
    \sum_{\lambda\in\Lambda_{\rm DG}}\Omega^\lambda_{\rm DG}=1
    \,.
   \label{eq:sum_DG}
\end{align}
From \eqref{eq:htilde_f}, \eqref{eq:2pt_spin2}, and \eqref{eq:omega_dg}, the two-point correlation function of the spin-2 dark matter signal in Fourier space is calculated as
\begin{align}
    \left<\tilde{h}^*_I(f)\tilde{h}_J(f')\right>
    &=\alpha^2f_{\rm DG}\calA\sum_{\lambda\in\Lambda_{\rm DG}}\Omega^\lambda_{\rm DG}\gamma^\lambda_{IJ}
    \nn
    &\times\delta\left(|f|-\frac{m}{2\pi}\right)\delta\left(f-f'\right)\,.
\label{eq:2pt_DG}
\end{align}
We have ignored the terms proportional to $\delta(f+f')$ because they vanish if we consider the randomness of $\theta$.

We have defined a dimensionless constant $\calA$ as
\begin{align}
    \calA\equiv\frac{\rho_{\rm DM}}{5\omega^2\mpl^2}\,,
\end{align}
The ratio of the energy density of the spin-2 ULDM to the total local density of dark matter $\rho_{\rm DM}$ is defined as
\begin{align}
    f_{\rm DG}\equiv\frac{\rho_{\rm DG}}{\rho_{\rm DM}}\,,
\end{align}
where $\rho_{\rm DM}\simeq 0.3\,{\rm GeV/cm^{3}}$ around the solar system in our galaxy. We have also defined the overlap reduction function (ORF) for each polarization mode by
\begin{align}
    \gamma^\lambda_{IJ}&\equiv 5\int\frac{d^2\hat{\bm{\Omega}}}{4\pi}F^\lambda_I(\hat{\bm{\Omega}}) F^\lambda_J(\hat{\bm{\Omega}})\,,
    \label{eq:spin-2ORF}
\end{align}
for $\lambda\in\Lambda_{\rm DG}$. They are normalized to be unity if $I=J$. These ORFs quantify the relative orientation and position of two detectors \cite{Flanagan:1993ix}. The extension for the vector and scalar polarization modes in extended gravitational waves are provided by \cite{Nishizawa:2009bf}. ORFs for the spin-2 dark matter correspond to the long wavelength limit of such extended gravitational waves.
By applying \eqref{eq:2pt_DG} to the formula \eqref{eq:SNR_fomula}, we obtain the following formula of SNR for the spin-2 dark matter
\begin{align}
    {\rm SNR}^{\rm DG}_{IJ}&=\Big|\frac{2\sqrt{2}\alpha^2 f_{\rm DG}\calA T_{\rm eff}}{\sqrt{S_{n,I}(\frac{m}{2\pi})S_{n,J}(\frac{m}{2\pi})}}
    \sum_{\lambda\in\Lambda_{\rm DG}}\Omega^\lambda_{\rm DG}\gamma^\lambda_{IJ}\Big|\,.\label{eq:snr_dg}
\end{align}

\subsection{Cross-correlation Search for Spin-1 ULDM}
The Fourier transformation of the signal  \eqref{eq:signal_dark_photon_h} is expressed as
\begin{align}
    \tilde{h}(f)=\tilde{h}_{\rm time}(f)+\tilde{h}_{\rm space}(f)\,,
\end{align}
where
\begin{align}
    \tilde{h}_{\rm time}(f)&=\frac{\epsilon_D e}{i\omega L}\frac{Q_D}{M}\sin^2\left(\frac{\omega L}{2}\right)
    \nn
    &\times\sum_{\lambda\in\Lambda_{\rm DP}}\int d^2\hat{\bm{\Omega}} A^\lambda (\hat{\bm{\Omega}}) G^\lambda(\hat{\bm{\Omega}})
    \nn
    &\times\left[\delta\left(f-\frac{\omega}{2\pi}\right)e^{i\theta}-\delta\left(f+\frac{\omega}{2\pi}\right)e^{-i\theta}\right]
    \,,
    \\
    \tilde{h}_{\rm space}(f)&=-\frac{1}{2}\epsilon_D e v \frac{Q_D}{M}
    \nn
    &\times\sum_{\lambda\in\Lambda_{\rm DP}}\int d^2\hat{\bm{\Omega}} A^\lambda(\hat{\bm{\Omega}}) F^\lambda(\hat{\bm{\Omega}})
    \nn
    &\times\left[\delta\left(f-\frac{\omega}{2\pi}\right)e^{i\omega L+i\theta}+\delta\left(f+\frac{\omega}{2\pi}\right)e^{-\i\omega L-i\theta}\right]
    \,.
\end{align}
Similar to the spin-2 ULDM case, the two-point function of the field amplitudes is expressed by
\begin{align}
    \left<A^\lambda(\hat{\bm{\Omega}})A^{\lambda'}(\hat{\bm{\Omega}}')\right>=H_{\rm DP}^\lambda \delta_{\lambda\lambda'}\delta^2\left(\hat{\bm{\Omega}},\hat{\bm{\Omega}}'\right)\,.
\end{align}
From this expression, the energy density of the spin-1 ULDM is given by
\begin{align}
    \rho_{\rm DP}=\frac{1}{2}\left<\dot{A}_i(t,\bm{x})\dot{A}_i(t,\bm{x})\right>=\pi \omega^2\sum_{\lambda\in\Lambda_{\rm DP}} H_{\rm DP}^\lambda\,.
\end{align}
Furthermore, the density parameters of each polarization mode is defined as
\begin{align}
    \Omega^\lambda_{\rm DP}=\frac{\pi \omega^2 H_{\rm DP}^\lambda}{\rho_{\rm DP}}\,,
\end{align}
for $\lambda\in\Lambda_{\rm DP}$, which satisfies
\begin{align}
    \sum_{\lambda\in\Lambda_{\rm DP}}\Omega^\lambda_{\rm DP}=1
    \,.
\end{align}

The cross-correlation can be calculated as the sum of the following two terms
\footnote{
The cross terms of $h_{\rm time}(f)$ and $h_{\rm space}(f)$ such as $\left<\tilde{h}_{{\rm space},I}^*\tilde{h}_{{\rm time},J}\right>$ and $\left<\tilde{h}_{{\rm time},I}^*\tilde{h}_{{\rm space},J}\right>$ are neglected because their ORF vanishes as shown in Appendix~\ref{sec:app2}. 
}
\begin{align}
    &\left<\tilde{h}_{{\rm time},I}(f)^*\tilde{h}_{{\rm time},J}(f')\right>
    \nn
    &=\epsilon_D^2 f_{\rm DP}\calB_{IJ}\sum_{\lambda\in\Lambda_{\rm DP}}\Omega^{\lambda}_{\rm DP}{\Gamma}^{\lambda}_{IJ}
    \delta\left(|f|-\frac{\omega}{2\pi}\right)\delta(f-f')\,,
    \label{eq:time_time}
    \\
    &\left<\tilde{h}_{{\rm space},I}(f)^*\tilde{h}_{{\rm space},J}(f')\right>
    \nn
    &=\epsilon_D^2 f_{\rm DP}\calC\left[\sum_{\lambda\in\Lambda_{\rm DP}}\Omega_{\rm DP}^\lambda \gamma^\lambda_{IJ}+\frac{1}{3}\Omega^{z}_{\rm DP}{\gamma}^{z}_{IJ}\right]
    \nn
    &\times\delta\left(|f|-\frac{\omega}{2\pi}\right)\delta(f-f')\,,
    \label{eq:space_space}
\end{align}
In these expressions, we have defined the dimensionless coefficients $\calB_{IJ}$ and $\calC$ as
\begin{align}
    \calB_{IJ}&\equiv\frac{8\rho_{\rm DM}e^2}{3\omega^4L_I L_J}\left(\frac{Q_D}{M}\right)^2
    \sin^2\left(\frac{\omega L_I}{2}\right)\sin^2\left(\frac{\omega L_J}{2}\right)\,,
    \\
    \calC&\equiv\frac{\rho_{\rm DM}e^2v^2}{5\omega^2}\left(\frac{Q_D}{M}\right)^2\,,
\end{align}
and the density fraction of the spin-1 ULDM to the density of the total dark matter as
\begin{align}
    f_{\rm DP}&\equiv\frac{\rho_{\rm DP}}{\rho_{\rm DM}}\,.
\end{align}
We have also defined two types of ORF. One is induced by the finite-time traveling effects as
\begin{align}
    {\Gamma}^{\lambda}_{IJ}&\equiv\frac{3}{2}\int\frac{d^2\hat{\bm{\Omega}}}{4\pi} G^\lambda_I(\hat{\bm{\Omega}}) G^\lambda_J(\hat{\bm{\Omega}})\,,
\end{align}
for $\lambda\in\Lambda_{\rm DP}$. Another ORF arises from the spatial displacement of the mirrors: 
\begin{align}
    \gamma^{x}_{IJ}&\equiv 5\int\frac{d^2\hat{\bm{\Omega}}}{4\pi}F^{x}_{I}(\hat{\bm{\Omega}})F^{x}_{J}(\hat{\bm{\Omega}})
    \,,
    \\
    \gamma^{y}_{IJ}&\equiv 5\int\frac{d^2\hat{\bm{\Omega}}}{4\pi}F^{y}_{I}(\hat{\bm{\Omega}})F^{y}_{J}(\hat{\bm{\Omega}})
    \,,
    \\
    \gamma^{z}_{IJ}&\equiv\frac{15}{4}\int\frac{d^2\hat{\bm{\Omega}}}{4\pi}F^z_{I}(\hat{\bm{\Omega}})F^z_{J}(\hat{\bm{\Omega}})\,.
    \label{eq:gamma_z}
\end{align}
Note that while we have $\gamma^x_{IJ}=\gamma_{IJ}^{V_x}$ and $\gamma^y_{IJ}=\gamma_{IJ}^{V_y}$, $\gamma^z_{IJ}$ differs from the ORF of the scalar polarization of spin-2 ULDM, denoted as $\gamma^S_{IJ}$. This discrepancy arises because $e_{ij}^S$ is composed of not only longitudinal mode but also breathing mode, as specified in \eqref{eq:esij}.

By substituting \eqref{eq:time_time} and \eqref{eq:space_space} into \eqref{eq:SNR_fomula}, the SNR for spin-1 ULDM is obtained as
\begin{align}
    {\rm SNR}^{\rm DP}_{IJ}&\approx\frac{2\sqrt{2}\epsilon_D^2 f_{\rm DP}T_{\rm eff}}{\sqrt{S_{n,I}(\frac{m}{2\pi})S_{n,J}(\frac{m}{2\pi})}}
    \Bigg|\calB_{IJ}\sum_{\lambda\in\Lambda_{\rm DP}}\Omega_{\rm DP}^\lambda\Gamma^\lambda_{IJ}
    \nn
    &+\calC\left[\sum_{\lambda\in\Lambda_{\rm DP}}\Omega_{\rm DP}^\lambda\gamma^\lambda_{IJ}+\frac{1}{3}\Omega_{\rm DP}^z\gamma_{IJ}^z\right]\Bigg|\,.\label{eq:snr_dp}
\end{align}

\subsection{Cross correlation Search for Spin-0 ULDM}
The SNR for spin-0 ULDM is calculated in the same manner as for the spin-1 ULDM. The cross-correlation for the spin-0 ULDM signal is given by the sum of the following two terms:
\begin{align}
    \left<h^*_{{\rm time},I}(f)h_{{\rm time},J}(f')\right>&= f_{\rm SD}\alpha_A^2\calB'_{IJ}\Gamma_{IJ}^z
    \nn
    &\times\delta\left(|f|-\frac{\omega}{2\pi}\right)\delta(f-f')\,,
    \label{eq:cross_spin0_time}
    \\
     \left<h^*_{{\rm space},I}(f)h_{{\rm space},J}(f')\right>&= f_{\rm SD}\alpha_A^2 \calC'_{IJ}\gamma_{IJ}^z 
     \nn
     &\times\delta\left(|f|-\frac{\omega}{2\pi}\right)\delta(f-f')\,,   
     \label{eq:cross_spin0_space}
\end{align}
where 
\begin{align}
    \calB'_{IJ}&=\frac{8\kappa^2\rho_{\rm DM}v^2}{3\omega^4 L_IL_J}\sin^2\left(\frac{\omega L_I}{2}\right)\sin^2\left(\frac{\omega L_J}{2}\right)\,,
    \\
    \calC'&=\frac{4\kappa^2v^4\rho_{\rm DM}}{15\omega^2}\,.
\end{align}
and
\begin{align}
    f_{\rm SD} \equiv\frac{\rho_{\rm SD}}{\rho_{\rm DM}}\,,
\end{align}
with $\rho_{\rm SD}$ being the energy density of the spin-0 ULDM:
\begin{align}
    \rho_{\rm SD}=\frac{1}{2}\left<\dot{\phi}(t,\bm{x})\dot{\phi}(t,\bm{x})\right>
    \,.
\end{align}
By substituting \eqref{eq:cross_spin0_time} and \eqref{eq:cross_spin0_space} into \eqref{eq:SNR_fomula}, we obtain
\begin{align}
    {\rm SNR}^{\rm SD}_{IJ}\approx\left|\frac{2\sqrt{2}f_{\rm SD}\alpha_A^2T_{\rm eff}(\calB'_{IJ}\Gamma^z_{IJ}+\calC'\gamma^z_{IJ})}{\sqrt{S_{n,I}(\frac{m}{2\pi})S_{n,J}(\frac{m}{2\pi})}}\right|\,.
    \label{eq:SNR_spin-0}
\end{align}

\section{Overlap reduction function and its implications}
\label{sec:ORF}

\subsection{Overlap reduction function}

\begin{figure}
    \centering
    \includegraphics[scale=0.3]{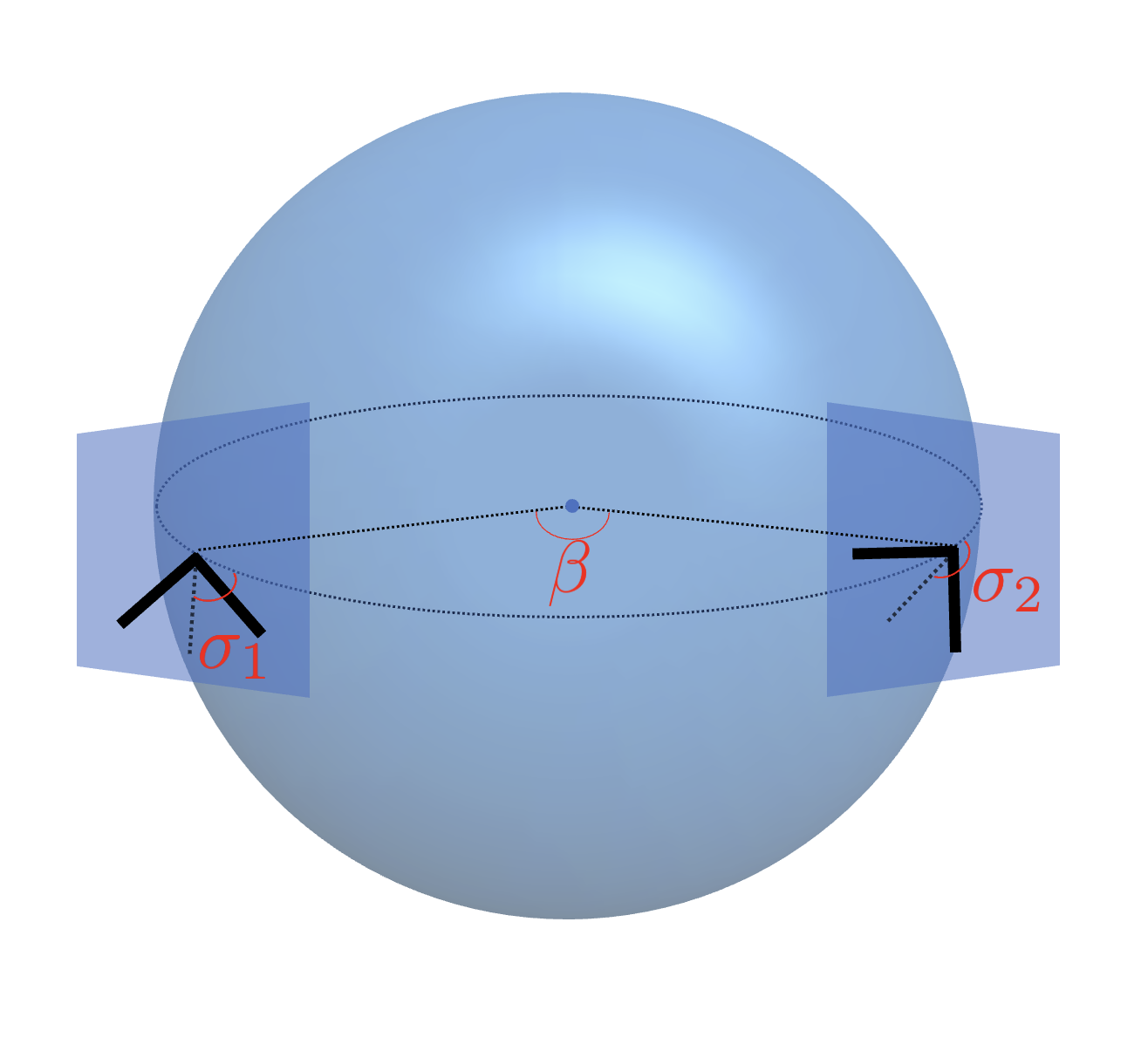}
    \caption{Illustration of  parameters $\left(\beta, \sigma_1, \sigma_2\right)$ on the Earth coordinate. L shaped bold lines are detector 1 and 2. $\sigma_1$ and $\sigma_2$ represent the angles between the great circles of two detectors and the bisectors of each detector, respectively. $\beta$ denotes the central angle between the two detectors.}
    \label{fig:angle_definition}
\end{figure}

\begin{table}
    \caption{Parameters $(\beta,\delta,\Delta,\gamma_{IJ},\Gamma_{IJ})$ for the various combinations of detectors in units of degrees~\cite{COLAB,Omiya:2023rhj}. In this table, I, K, H, and L represent LIGO-India, KAGRA, LIGO-Hanford, and LIGO-Livingston, respectively. }
    
\begin{tabular}{cccccc}
    \hline\hline
       Detectors & $\beta$ & $\delta$ & $\Delta$ & $\gamma_{IJ}$ & $\Gamma_{IJ}$ \\\hline\hline
    KL & 54.8912 & 28.5968  & 102.703  & -0.280        & 0.620  \\
    IH & 112.279 & -79.7392 & 164.518  & -0.150        & -0.880 \\
    IL & 128.472 & 35.8711  & 203.311  & 0.009         & -0.500 \\
    IV & 59.7883 & 42.8974  & 159.522  & -0.570        & -0.130 \\
    KH & 72.3721 & -88.2092 & 117.523  & 0.460         & -0.450 \\
    KL & 99.2727 & -135.139 & 160.223  & -0.240        & -0.450 \\
    KV & 86.5230  & 31.6663  & 97.443   & -0.360        & 0.690  \\
    HL & 27.2233 & -44.9735 & 241.550   & -0.890        & 0.031  \\
    HV & 79.6176 & -28.856  & 144.217  & -0.015        & 0.190  \\
    LV & 76.7637 & 27.0677  & 172.467  & -0.250        & -0.012 \\    
    \hline\hline
\end{tabular}
\label{tab:BETA_SIGMA_GAMMA_2}
\end{table}

Considering two different ground-based detectors, designated as $I$ and $J$, their orientation and relative position are determined by three parameters, $\sigma_I, \sigma_J$, and $\beta$, as depicted in Fig.~\ref{fig:angle_definition}. We then introduce two combinations of angles, 
\begin{align}
    \Delta \equiv \frac{\sigma_I+\sigma_J}{2}, \quad \delta \equiv \frac{\sigma_I-\sigma_J}{2}\,.
    \label{eq:notation_sigma}
\end{align}
Additionally, we define $\hat{\bm{d}}$, which represents a unit vector along the direction from detector $J$ to $I$.

Conventionally, in the context of polarization search in gravitational waves, the ORFs for tensor, vector, and scalar polarizations are defined as~\cite{Flanagan:1993ix,Nishizawa:2009bf} 
\begin{align}
    \gamma_{IJ}^T&\equiv\frac{1}{2}(\gamma_{IJ}^+ +\gamma_{IJ}^\times)\,,
    \label{gammaT}
    \\
    \gamma_{IJ}^V&\equiv\frac{1}{2}(\gamma_{IJ}^{V_x} +\gamma_{IJ}^{V_y})\,,
    \\
    \gamma_{IJ}^{S'}&\equiv\frac{3}{1+2q}(\gamma_{IJ}^b+q\gamma_{IJ}^\ell)\,,
    \label{gammaS'}
\end{align}
where $\gamma_{IJ}^b$ and $\gamma_{IJ}^\ell$ are the ORF of breathing and longitudinal polarization defined in the similar way as \eqref{eq:spin-2ORF}. For the spin-2 dark matter signal, the scalar polarization is given by a specific linear combination of breathing and longitudinal polarization with $q=\sqrt{2}$ for which $\gamma_{IJ}^{S'}$ reduces to $\gamma_{IJ}^S$ defined by \eqref{eq:spin-2ORF}. The ORF for the spin-2 ULDM corresponds to the long wavelength limit of the ORF \eqref{gammaT}-\eqref{gammaS'}. Thus, with the parameters $(\beta,\Delta,\delta)$, ORFs for spin-2 ULDM are expressed as~\cite{Flanagan:1993ix}
\begin{align}
    \gamma_{IJ}\left(\beta, \Delta,\delta\right)&\equiv\gamma_{IJ}^T=\gamma_{IJ}^V=\gamma_{IJ}^S
    \label{eq:same_ORF_sipn-2}
    \\
    &=  \cos^4\left(\frac{\beta}{2}\right)\cos(4\delta)-\sin^4\left(\frac{\beta}{2}\right)\cos(4\Delta)\,.
    \label{eq:ORF_spin-2}
\end{align}

As for the spin-1 and spin-0 ULDM signals, there are two different contributions: one is from the spatial displacement of the mirrors $\gamma_{IJ}^{\lambda}$ and the other is from the finite-time traveling effect $\Gamma_{IJ}^{\lambda}$.
The former ones are defined by
\begin{align}
    \gamma_{IJ}^{\mathsf{T}}&\equiv\frac{1}{2}(\gamma_{IJ}^{x} +\gamma_{IJ}^{y})=\gamma^V_{IJ}\,,
    \\
    \gamma_{IJ}^{\mathsf{L}}&\equiv\gamma^z_{IJ}\,.
\end{align}
They reduce to the identical form to \eqref{eq:ORF_spin-2}:
\begin{align}
    \gamma_{IJ}&=\gamma_{IJ}^\mathsf{T}=\gamma_{IJ}^\mathsf{L}\nn
    &=\cos^4\left(\frac{\beta}{2}\right)\cos(4\delta)-\sin^4\left(\frac{\beta}{2}\right)\cos(4\Delta)\,.
\end{align}
For the finite-time traveling effect, ORFs for the transverse and longitudinal modes are defined as
\begin{align}
    \Gamma_{IJ}^{\mathsf{T}}&\equiv\frac{1}{2}(\Gamma_{IJ}^x +\Gamma_{IJ}^y)\,,
    \\
    \Gamma_{IJ}^{\mathsf{L}}&\equiv\Gamma^z_{IJ}\,.\label{eq:GammaTLdef}
\end{align}
As shown in Appendix~\ref{sec:tildeORF}, they are rewritten as 
\begin{align}
    \Gamma_{IJ}\left(\beta, \Delta,\delta\right)&\equiv\Gamma_{IJ}^{\mathsf{T}}=\Gamma_{IJ}^{\mathsf{L}}
    \label{eq:same_ORF_sipn-1}
    \\
    &=\cos^2\left(\frac{\beta}{2}\right)\cos(2\delta)
    -\sin^2\left(\frac{\beta}{2}\right)\cos(2\Delta)\,.
    \label{eq:limGamma}
\end{align}
As shown in \eqref{eq:ORF_spin-2} and \eqref{eq:limGamma}, the dependency of ORFs for $(\beta,\Delta,\delta)$ differs, reflecting the symmetry of the detector tensor/vector.

The SNRs for ULDM of spin-2 \eqref{eq:snr_dg}, spin-1 \eqref{eq:snr_dp}, and spin-0 \eqref{eq:SNR_spin-0} respectively reduce to
\begin{align}
    {\rm SNR}^{\rm DG}_{IJ}&\approx\left|\frac{2\sqrt{2}\alpha^2 f_{\rm DG}T_{\rm eff}\calA\gamma_{IJ}}{\sqrt{S_{n,I}(\frac{m}{2\pi})S_{n,J}(\frac{m}{2\pi})}}\right|\,,
    \label{eq:SNR_spin2DM}
    \\
    {\rm SNR}^{\rm DP}_{IJ}&\approx\left|\frac{2\sqrt{2}\epsilon_D^2 f_{\rm DP}T_{\rm eff}(\calB_{IJ}\Gamma_{IJ}+\calC(1+\frac{1}{3}\Omega^z_{\rm DP})\gamma_{IJ})}{\sqrt{S_{n,I}(\frac{m}{2\pi})S_{n,J}(\frac{m}{2\pi})}}\right|\,,
    \label{eq:SNR_DP}
     \\
    {\rm SNR}^{\rm SD}_{IJ}&\approx\left|\frac{2\sqrt{2}\alpha_A^{2} f_{\rm SD}T_{\rm eff}(\calB'_{IJ}\Gamma_{IJ}+\calC'\gamma_{IJ})}{\sqrt{S_{n,I}(\frac{m}{2\pi})S_{n,J}(\frac{m}{2\pi})}}\right|
    \,.
    \label{eq:SNR_SD}
\end{align}
For the spin-2 ULDM signal, the SNR is independent of how the energy density is assigned to each polarization mode, while the SNR of the spin-1 ULDM signal depends on $\Omega^z_{\rm DP}$. For the purpose of our analysis, we assume $\Omega_{\rm DP}^z=1/3$ in the following. This assumption corresponds to the randomized polarization scenario.

We proceed to evaluate the ORFs for the cross-correlation of five ground-based detectors: LIGO-Livingston (L), LIGO-Hanford (H), LIGO-India (I), Virgo (V), and KAGRA (K). Table \ref{tab:BETA_SIGMA_GAMMA_2} presents the values of $(\beta,\Delta,\delta)$ for each combination of detectors. Additionally, Table \ref{tab:BETA_SIGMA_GAMMA_2} displays the ORFs, $(\gamma_{IJ},\Gamma_{IJ})$, computed using \eqref{eq:ORF_spin-2}, \eqref{eq:limGamma}, and Table~\ref{tab:BETA_SIGMA_GAMMA_2}. Interestingly, for the LIGO-Livingston and Hanford pair, $|\Gamma_{\rm HL}|$ is much smaller than $|\gamma_{\rm HL}|$. This means a substantial suppression of SNR, which consequently influences the current constraint of spin-1 ULDM with gravitational wave detectors. This issue will be further addressed in the next subsection.

\subsection{Current Constraints}

\begin{figure*}
    \begin{minipage}{0.5\hsize}
         \includegraphics[width=\linewidth]{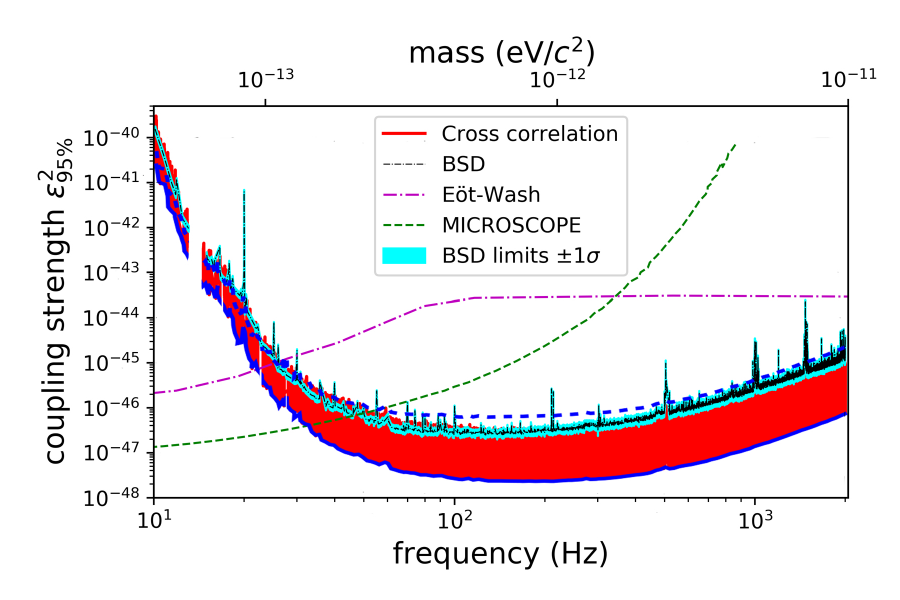}
    \end{minipage}
    \begin{minipage}{0.45\hsize}
         \includegraphics[width=\linewidth]{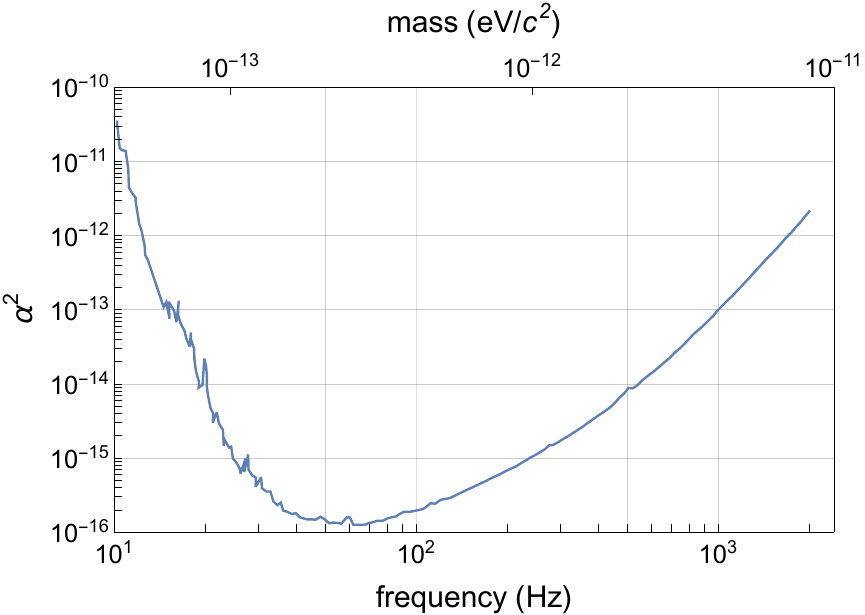}
    \end{minipage}
    \caption{(Left panel) The upper bound of the coupling constant $\epsilon_B^2$ for spin-1 ULDMs to standard model particles, derived from LIGO's and Virgo's O3 run data. This figure is adapted from \cite{LIGOScientific:2021ffg}. The red line represents the constraint obtained from the cross-correlation between LIGO-Hanford and Livingston. While this analysis incorporates the finite-time traveling effect, it employed the ORF for the standard gravitational waves. We added two blue lines. Due to the change of the ORF for the finite-time traveling effect, the lower bound (blue solid line) should be corrected to the blue dashed line.
    (Right panel) The upper bound of the coupling constant $\alpha^2$ for spin-2 ULDM to standard model particles. These constraints are obtained by scaling the results from the O3 data related to the upper limit of the coupling constant for spin-1 ULDM. For both plots, $f_{\rm DP}=f_{\rm DG}=1$ is assumed.}
    \label{fig:LVK_fig}
\end{figure*}

\subsubsection{Current Constraints on Spin-1 ULDM}

For spin-1 ULDM, ref.~\cite{LIGOScientific:2021ffg} updated the constraint on the coupling constant $\epsilon_D$ through a correlation analysis between LIGO-Livingston and Hanford, utilizing data from LIGO's O3 run. While this study did take the finite-time traveling effect into account, it uses the ORF defined for the standard gravitational waves. However, as we pointed out, the ORF for the finite-time traveling effect is different from the standard one.
Since $|\Gamma_{\rm HL}|$ is about 0.035 times  smaller than $|\gamma_{\rm HL}|$, the constraint is expected to be weaker when the finite-time traveling effect gives the dominant contribution to the SNR.

Figure~\ref{fig:LVK_fig} is adapted from the figure illustrating the upper limit on $\epsilon_B^2$ in Ref.~\cite{LIGOScientific:2021ffg}. The red line represents the limit derived from the correlation analysis, without considering the modification of the ORF. We have added two blue lines. The lower boundary of the red region, depicted by the solid blue line, is expected to be modified as indicated by the blue dashed line. This modification is achieved by scaling the blue solid line with
\begin{align}
    \frac{(\calB_{\rm HL}+\frac{10}{9}\calC)\gamma_{\rm HL}}{\calB_{\rm HL}\Gamma_{\rm HL}+\frac{10}{9}\calC\gamma_{\rm HL}}\,,
\end{align}
where we have set $\Omega_{\rm DP}^z=1/3$. As a result, the current upper bound of $\epsilon_B^2$ is weakened by around 30 times.

\subsubsection{Current Constraint on Spin-2 ULDM}

By scaling the solid blue line in the left panel of Fig.~\ref{fig:LVK_fig} with
\begin{align}
    \frac{(\calB_{\rm HL}+\frac{10}{9}\calC)\gamma_{\rm HL}}{\calA \gamma_{\rm HL}}\,,
\end{align}
we can estimate the upper bound of the coupling constant $\alpha^2$ of spin-2 dark matter to standard model particles, as shown in the right panel of Fig.~\ref{fig:LVK_fig}. When spin-2 ULDM comprises a part of dark matter, the constraint on $\alpha^2$ is replaced by a constraint on $f_{\rm DG}\alpha^2$.

\subsection{Future constraints}

\begin{figure*}
    \begin{tabular}{cc}
    \begin{minipage}{0.45\hsize}
         \includegraphics[width=\linewidth]{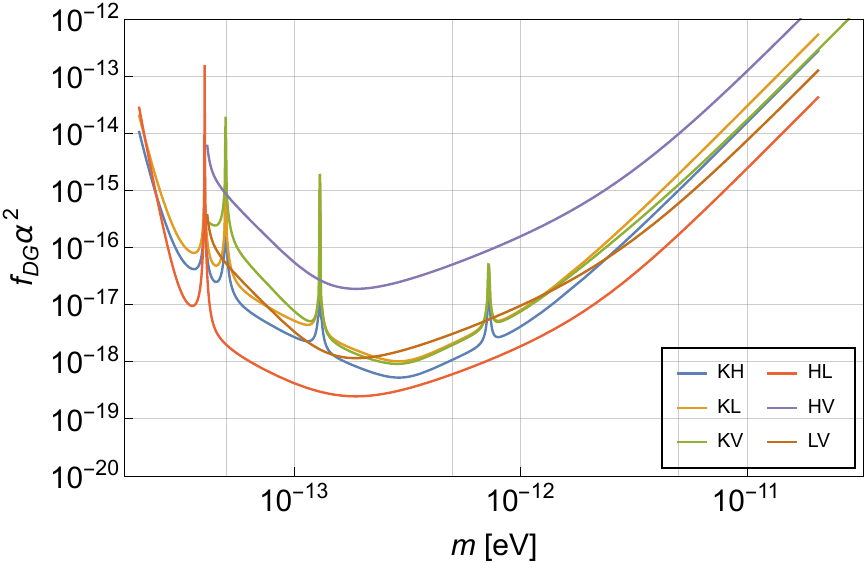}
    \end{minipage}
    \begin{minipage}{0.45\hsize}
         \includegraphics[width=\linewidth]{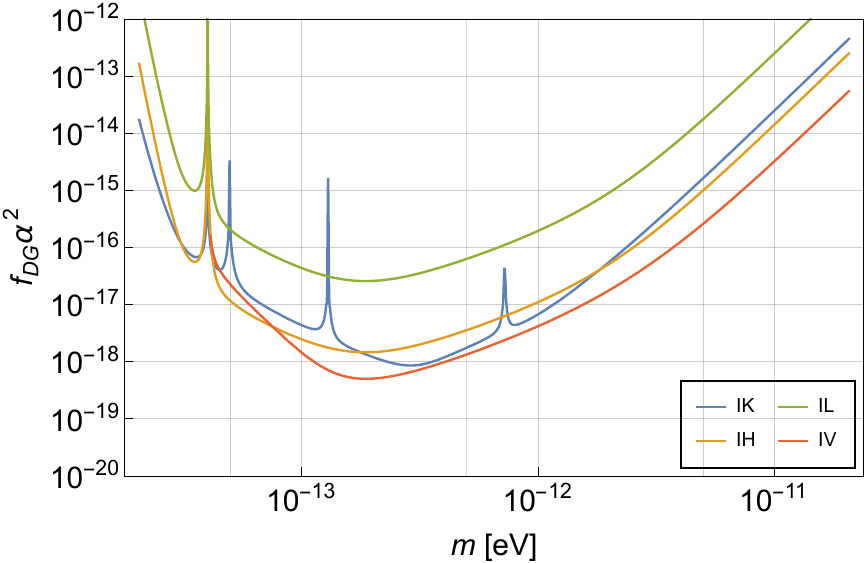}
    \end{minipage}
    \\
    \begin{minipage}{0.45\hsize}
         \includegraphics[width=\linewidth]{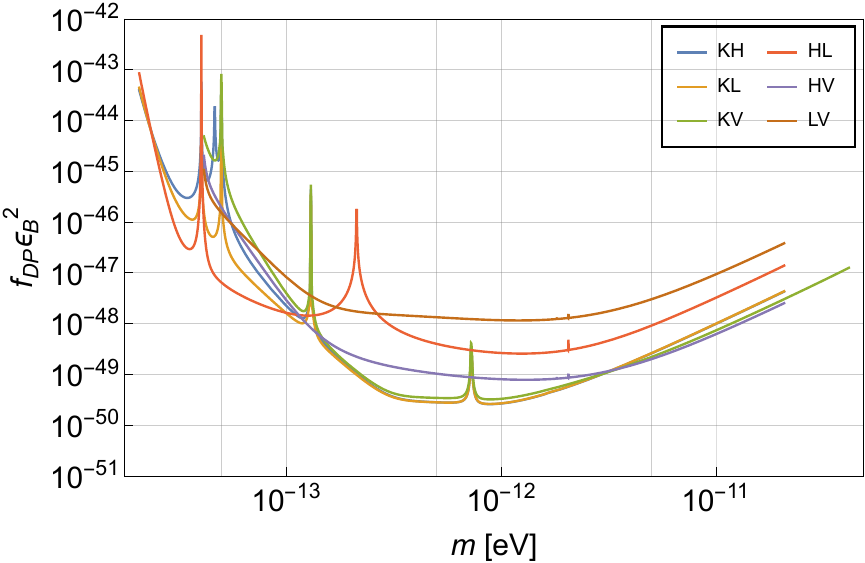}
    \end{minipage}
    \begin{minipage}{0.45\hsize}
         \includegraphics[width=\linewidth]{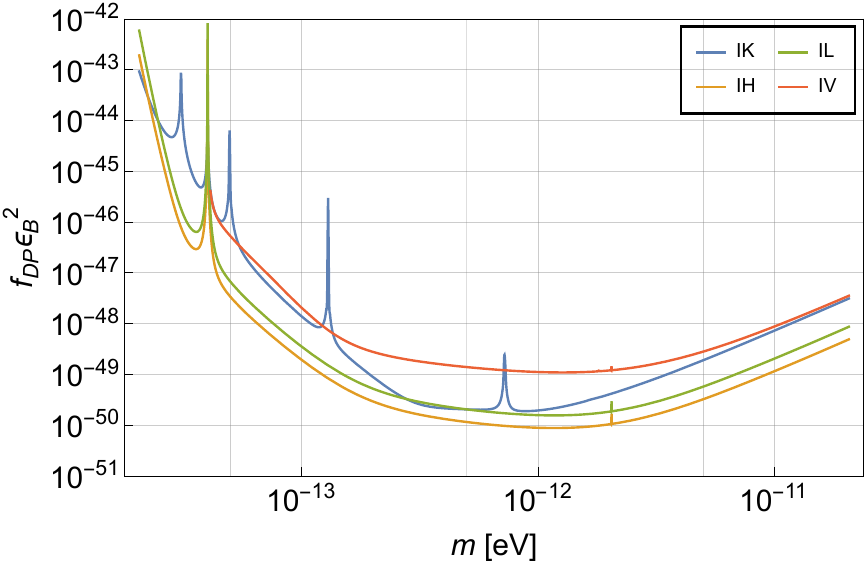}
    \end{minipage}
    \\
    \begin{minipage}{0.45\hsize}
         \includegraphics[width=\linewidth]{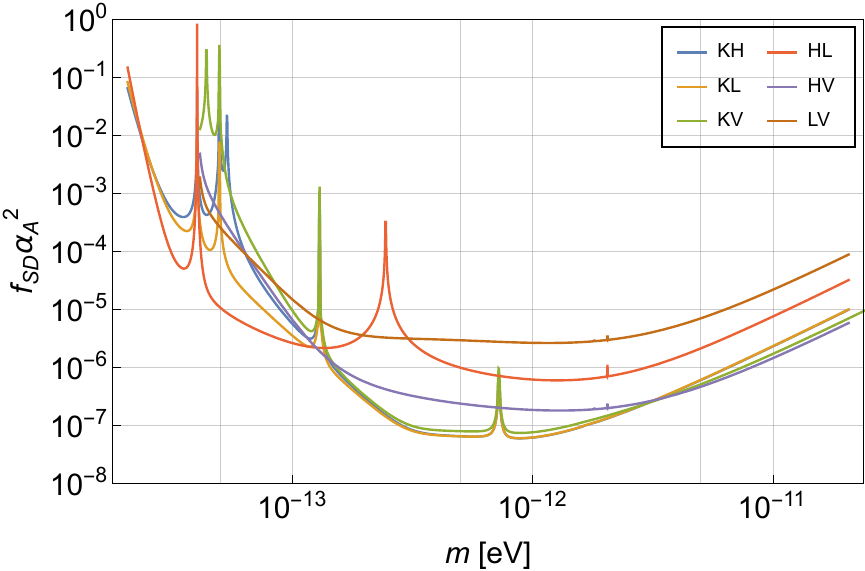}
    \end{minipage}
    \begin{minipage}{0.45\hsize}
         \includegraphics[width=\linewidth]{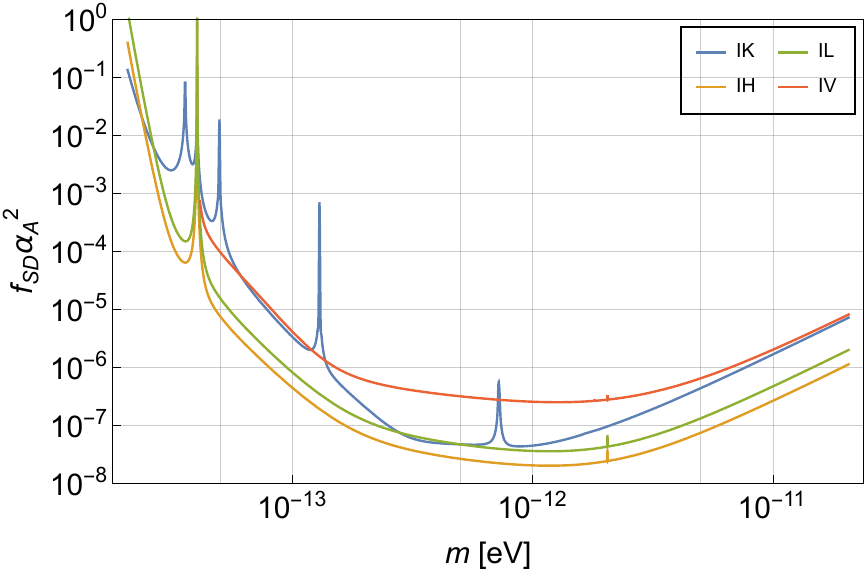}
    \end{minipage}
    \end{tabular}
    \caption{Expected upper bound for the coupling constants of spin-2 (upper), spin-1 (middle), and spin-0 (lower) ULDM to the standard model particles in cross-correlation search with LIGO-Livingston (L), LIGO-Hanford (H), LIGO-India (I), Virgo (V), and KAGRA (K). We assumed that ${\rm SNR}=1,T_{\rm obs}=1{\rm yr}.$ Noise power spectra are adapted from~\cite{JGW-T1707038,LIGO-T1800044,LIGO-P1200087}.}
    \label{fig:future_constraints}
\end{figure*}

We estimate the anticipated constraints on the coupling constants of spin-1 and spin-2 ULDM using a cross-correlation analysis. This analysis is based on the projected sensitivities of ground-based gravitational wave detectors, namely LIGO-Livingston (L), LIGO-Hanford (H), LIGO-India (I), Virgo (V), and KAGRA (K).

Figure \ref{fig:future_constraints} illustrates these constraints on the upper limits of the coupling constants for spin-2 (upper panel), spin-1 (middle panel), and spin-0 (lower panel) ULDM. They are derived from equations \eqref{eq:SNR_spin2DM}, \eqref{eq:SNR_DP}, and \eqref{eq:SNR_SD}, respectively. We make the assumption that ${\rm SNR}=1$ and $T_{\rm obs}=1{\rm yr}$. We have utilized publicly available data for noise power spectra in our calculations \cite{JGW-T1707038,LIGO-T1800044,LIGO-P1200087}. The left panel of Figure \ref{fig:future_constraints} presents the expected constraints when combining data from LIGO-Livingston (L), LIGO-Hanford (H), Virgo (V), and KAGRA (K). The right panel demonstrates the expected constraints when LIGO-India (I) is also included.

\section{Distinguishing Spin of ULDM}
\label{sec:ExploringSpinofULDM}

\begin{figure*}
    \begin{tabular}{cc}
    \begin{minipage}[t]{0.45\hsize}
    \includegraphics[width=\linewidth]{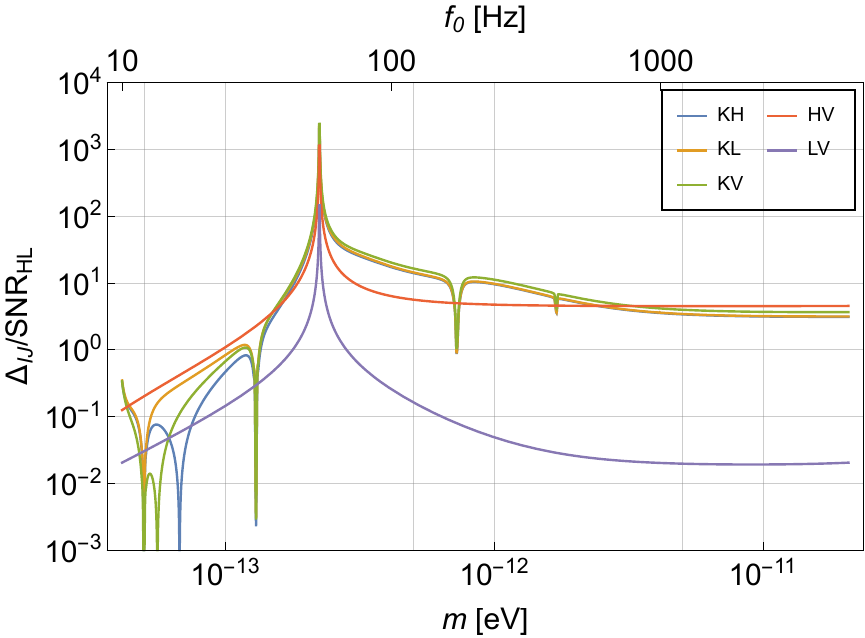}
    \end{minipage}
    \begin{minipage}[t]{0.45\hsize}
    \includegraphics[width=\linewidth]{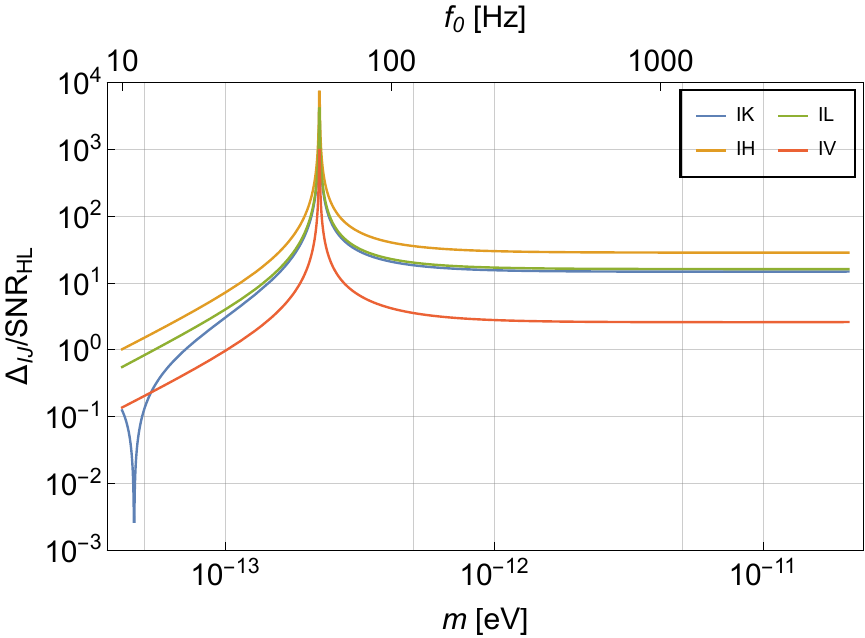}
    \end{minipage}
    \\
    \begin{minipage}[t]{0.45\hsize}
    \includegraphics[width=\linewidth]{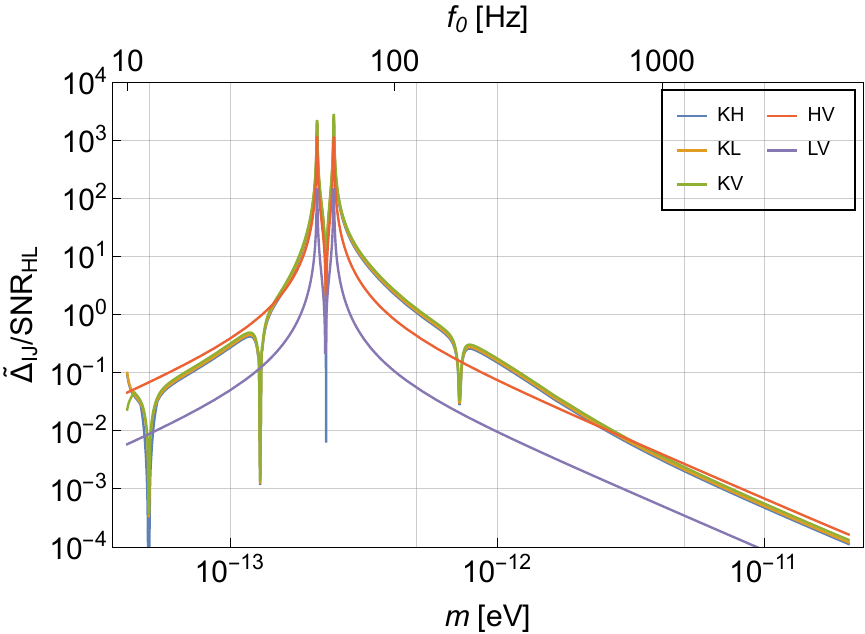}
    \end{minipage}
    \begin{minipage}[t]{0.45\hsize}
    \includegraphics[width=\linewidth]{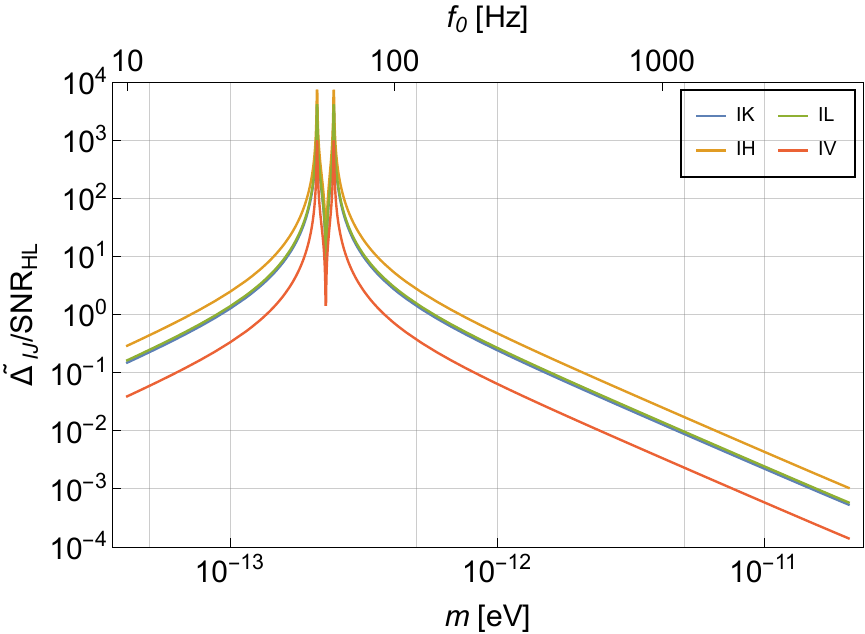}
    \end{minipage}
    \end{tabular}
    \caption{The distinguishability between the spin-2 and spin-1 ULDM (upper) and the spin-1 and spin-0 ULDM (lower) in the cross-correlation. $f_0\equiv m/2\pi$ represents the frequency of the signal. Note that the steep peaks are caused by the cancellation between $\calB_{IJ}\Gamma_{IJ}$ and $\calC\gamma_{IJ}$ in \eqref{Delta/SNR} (or $\calB_{IJ}'\Gamma_{IJ}$ and $\calC'\gamma_{IJ}$ in \eqref{eq:tildeSNR}) at which the spin-1 (or spin-0) UDLM has no sensitivity for a specific combination of detectors.}
    \label{fig:ratio}
\end{figure*}

\subsection{Distinguishing Signal Between Spin-2 and Spin-1 ULDM}

We explore the distinguishability between signals from spin-2 and spin-1 ULDM. Recall that the SNR for spin-2 is given by the standard ORF $\gamma_{IJ}$ while that for spin-1 is determined by $\gamma_{IJ}$ and $\Gamma_{IJ}$ where $\Gamma_{IJ}$ is the ORF from the finite-time traveling effect. As shown in \eqref{eq:ORF_spin-2} and \eqref{eq:limGamma}, these ORFs have different dependencies on the positions and orientations of the detectors (see also Table~\ref{tab:BETA_SIGMA_GAMMA_2}). In order to quantify these differences, we introduce a measure for the deviation in the SNR of cross-correlation between spin-2 and spin-1 ULDM signals as
\begin{align}
    \Delta_{IJ}\equiv\left|{\rm SNR}^{\rm DG}_{IJ}-{\rm SNR}^{\rm DP}_{IJ}\right|\,.\label{eq:delta_def}
\end{align}

As a case study, suppose that a ULDM signal is detected through the cross-correlation between LIGO-Livingston and Hanford with a certain SNR denoted by ${\rm SNR}_{\rm HL}$. The corresponding coupling constants are determined to be
\begin{align}
    \alpha^2 f_{\rm DG} &= \frac{{\rm SNR}_{\rm HL}\sqrt{S_{n,{\rm H}}(\frac{m}{2\pi})S_{n,{\rm L}}(\frac{m}{2\pi})}}{2\sqrt{2}|\calA \gamma_{\rm HL}|T_{\rm eff}}\,,
    \\
    \epsilon_D^2 f_{\rm DP}&=\frac{{\rm SNR}_{\rm HL}\sqrt{S_{n,{\rm H}}(\frac{m}{2\pi})S_{n,{\rm L}}(\frac{m}{2\pi})}}{2\sqrt{2}|\calB_{\rm HL}\Gamma_{\rm HL}+\frac{10}{9}\calC\gamma_{\rm HL}|T_{\rm eff}}\,,
\end{align}
where we have set $\Omega_{\rm DP}^z=\frac{1}{3}$ for simplicity. From this, $\Delta_{IJ}$ can be expressed as
\begin{align}
    \frac{\Delta_{IJ}}{{\rm SNR}_{\rm HL}}&=\left|\left|\frac{\gamma_{IJ}}{\gamma_{\rm HL}}\right|-\left|\frac{\calB_{IJ}\Gamma_{IJ}+\frac{10}{9} \calC\gamma_{IJ}}{\calB_{\rm HL}\Gamma_{\rm HL}+\frac{10}{9}\calC\gamma_{\rm HL}}\right|\right|
    \sqrt{\frac{S_{n,{\rm H}}S_{n,{\rm L}}}{S_{n,{I}}S_{n,{J}}}}\,.
    \label{Delta/SNR}
\end{align}
We have assumed the same $T_{\rm eff}$ for all combinations of detectors.

Figure \ref{fig:ratio} illustrates $\Delta_{IJ}$ normalized by ${\rm SNR}_{\rm HL}$. In lower mass ranges ($m\lesssim v/L \simeq 5\times 10^{-13}{\rm eV}$), the signal from the spatial displacement of mirrors dominates over the signal from the finite-time traveling effect in the spin-1 ULDM signal. This leads to a suppression in the distinguishability of spin, as the ORF for spin-1 ULDM resembles that of spin-2 ULDM. Since $\calB_{IJ}/\calC$ is approximately proportional to $m^2$, $\Delta_{IJ}$ grows with mass in this region. In contrast, in higher mass ranges ($m\gtrsim v/L \simeq 5\times 10^{-13}{\rm eV}$), the finite-time traveling effect becomes the dominant factor in the spin-1 ULDM signal, enhancing the distinguishability of spins.

Note that the cross-correlation between LIGO-Livingston and Hanford lacks sensitivity to the spin-1 ULDM at $m \simeq 5\times10^{13}{\rm eV}$, due to the cancellation of terms $\calB_{\rm HL}\Gamma_{\rm HL}$ and $\frac{10}{9}\calC\gamma_{\rm HL}$. This specific cancellation leads to a divergence in $\Delta_{IJ}/{\rm SNR}_{\rm HL}$. Likewise, similar cancellations occur in other combinations of detectors, resulting in $\Delta_{IJ}/{\rm SNR}_{\rm HL}=0$.

\subsection{Distinguishing Signal Between Spin-1 and Spin-0 ULDM}

Both signals from spin-1 and spin-0 ULDM are composed of the sum of $h_{\rm time}$ and $h_{\rm space}$. However, their ratios differ unless $\Omega_{\rm DP}^z = 1$. This suggests that signals from spin-1 and spin-0 ULDM can be distinguished only if the ULDM mass is close to the region at which $h_{\rm time}$ and $h_{\rm space}$ contribute equally.

Similarly to \eqref{eq:delta_def}, we define $\tilde{\Delta}_{IJ}$ to indicate the distinguishability of spin-0 and spin-1 ULDM as
\begin{align}
    \tilde{\Delta}_{IJ}=|{\rm SNR}_{IJ}^{\rm DP}-{\rm SNR}_{IJ}^{\rm SD}|\,.\label{eq:tilde_delta_def}
\end{align}
We again consider the case where LIGO-Hanford and LIGO-Livingston detect a ULDM signal with a certain SNR through their cross-correlation analysis. Assuming that this signal originates from spin-0 ULDM, $\alpha_A^2 f_{\rm SD}$ is determined by
\begin{align}
    \alpha_A^2 f_{\rm SD}&=\frac{{\rm SNR}_{\rm HL}\sqrt{S_{n,{\rm H}}(\frac{m}{2\pi})S_{n,{\rm L}}(\frac{m}{2\pi})}}{2\sqrt{2}|\calB'_{\rm HL}\Gamma_{\rm HL}+\calC'\gamma_{\rm HL}|T_{\rm eff}}\,.
\end{align}
Thus, $\tilde{\Delta}_{IJ}$ is rewritten by
\begin{align}
    \frac{\tilde{\Delta}_{IJ}}{{\rm SNR}_{\rm HL}}&=\left|\left|\frac{\calB_{IJ}\Gamma_{IJ}+(1+\frac{\Omega_{\rm DP}^z}{3})\calC\gamma_{IJ}}{\calB_{\rm HL}\Gamma_{\rm HL}+(1+\frac{\Omega_{\rm DP}^z}{3})\calC\gamma_{\rm HL}}\right|\right.
    \nn
    &\left.-\left|\frac{\calB'_{IJ}\Gamma_{IJ}+\calC'\gamma_{IJ}}{\calB'_{\rm HL}\Gamma_{\rm HL}+\calC'\gamma_{\rm HL}}\right|\right|
    \sqrt{\frac{S_{n,{\rm H}}S_{n,{\rm L}}}{S_{n,{I}}S_{n,{J}}}}\,.
    \label{eq:tildeSNR}
\end{align}
Since $\frac{\calB_{IJ}}{\calC}=\frac4 3\frac{\calB'_{IJ}}{\calC'}$ for the any combination of the detectors, \eqref{eq:tildeSNR} vanished in the limit of $\Omega_{\rm DP}^z=1$, it corresponds to the case that only longitudinal mode is excited.

The lower panels of Fig.~\ref{fig:ratio} show $\tilde{\Delta}_{IJ}$ normalized by ${\rm SNR}_{\rm HL}$. We have set $\Omega_{\rm DP}^z=\frac{1}{3}$. These plots have two peaks around $f\sim 50$ Hz that are caused by the cancellation of each denominator in \eqref{eq:tildeSNR}. Apart from the mass region around $m\sim v/L$, the distinguishability is damped because the same effect is dominated for both ULDM signals, and there are only a few difference in the position and orientation dependence on the ORFs.

\subsection{Maximal Likelihood}

In a practical ULDM search, the method of likelihood is useful for detecting the ULDM signal. We only consider the likelihood analysis to distinguish the signal between spin-1 and spin-2 ULDM, but the same analysis can be applied for distinguishing the spin-0 ULDM from others. The likelihood for each ULDM is given by
\begin{align}
    \calL(S|\mu^M)&=\prod_{IJ\in\calD}\frac{1}{\sqrt{2\pi\sigma^2_{IJ}}}
    \nn
    &\times\exp\left[-\frac{(S_{IJ}-\mu^M_{IJ}(\theta_M))^2}{2\sigma_{IJ}^2}\right]\,,
\end{align}
for $M=\{{\rm DG},{\rm DP}\}$. Here, $\calD$ denotes a set of detectors, and $S$ is a cross-correlation signal defined by \eqref{eq:cross-correlationalsignal}. The model parameters $\{\theta_{\rm DG},\theta_{\rm DP}\}=\{f_{\rm DG}\alpha^2,f_{\rm DP}\epsilon_D^2\}$ are decided by maximizing the likelihood for each ULDM. $\mu^M_{IJ}$ represents the mean value of the signal, which is given by \eqref{eq:mu_calc}. More explicitly, for each ULDM model, $\mu^M_{IH}$ are given by
\begin{align}
    \mu_{IJ}^{\rm DG}(\theta_{\rm DG})&\approx T_{\rm eff}\frac{\theta_{\rm DG}\calA|\gamma_{IJ}|}{\sqrt{S_{n,I}S_{n,J}}}\,,
    \\
    \mu_{IJ}^{\rm DP}(\theta_{\rm DP})&\approx T_{\rm eff}\frac{\theta_{\rm DP}|\calB_{IJ}\Gamma_{IJ}+(1+\frac{\Omega_{\rm DP}^z}{3})\calC\gamma_{IJ}|}{\sqrt{S_{n,I}S_{n,J}}}\,.
\end{align}
Therefore, the deviation of ORFs expressed in \eqref{eq:ORF_spin-2} and \eqref{eq:limGamma} are reflected in the likelihood through $\mu^M_{IJ}$. 

One of the indicators to distinguish the spin-1 and spin-2 ULDM is the ratio of the maximized likelihoods
\begin{align}
    \Lambda=\frac{{\rm sup}_{\theta_{\rm DG}}\left\{\calL(S|\mu^{\rm DG})\right\}}{{\rm sup}_{\theta_{\rm DP}}\left\{\calL(S|\mu^{\rm DP})\right\}}\,.
\end{align}
Roughly speaking, if $\Lambda\gg1$, the signal can be regarded as originating from the spin-2 ULDM whereas if $\Lambda\ll1$, the signal can be regarded as originating from spin-1 ULDM. This ratio corresponds to the Bayesian factor with the same prior probability for each dark matter model.

\section{Summary and discussion}
\label{sec:summary}

In this paper, we have addressed the distinguishability of  ULDM signals with different spins with the gravitational wave detectors.
First, we have discussed how the oscillations of ULDM cause a motion of mirrors of the gravitational wave detectors.
For the spin-0 and spin-1 UDLM, the finite-time traveling effects are critical for the ULDM search in ground-based detectors, while for spin-2 dark matter, even in its vector and scalar polarization modes, finite-time traveling effects give a negligible contribution. This reflects the nature that the spin-2 ULDM exerts differential motion on the two mirrors, whereas the spin-1 and spin-0 ULDM exert common motion on them. Next, we pointed out that the finite-time traveling effects give rise to the different ORF from the regular gravitational waves. 
As a result, the spin-2 ULDM and the lower spin ULDM yield different signals in the cross-correlation analysis.
As shown in Figure~\ref{fig:ratio}, if the ULDM signal is detected at a certain SNR by at least three detectors, it is possible to distinguish the spin-2 ULDM from that of spin-0 or 1.

Furthermore, we have found that the current constraint on the spin-1 ULDM becomes about 30 times weaker than that reported in Ref.~\cite{LIGOScientific:2021ffg} when we use the appropriate ORF, as shown in Fig.~\ref{fig:LVK_fig}. This is due to the fact that the ORF of the finite-time traveling effect between LIGO-Livingston (L) and LIGO-Hanford (H) is 0.035 times smaller than the standard ORF for their geometrical configurations. The simple solution to improve the constraints on the spin-1 ULDM would be to additionally utilize the data of Virgo (V) or KAGRA (K). At $m\sim 10^{-12}$ eV, the cross-correlations of KV and LV expected to give better constraints than that of HL.

While we have considered multiple detectors, the relative position and orientation between a detector and dark matter change in time thanks to the rotation of the Earth. Within the coherent time $\tau_{\rm coh}\sim 9300\,{\rm s}\,(10^{-12}{\rm eV}/m)$,
we may analyze correlations between the data in different time intervals of the same detector and observe the changes in the ORF over time. From \eqref{eq:ORF_spin-2} and \eqref{eq:limGamma}, $\gamma_{IJ}$ and $\Gamma_{IJ}$ are expected to have different time dependencies. Therefore, it may be beneficial to use different time data to distinguish between spin-1 and spin-2 ULDM based on the time dependence of the ORF. Since this method does not require multiple detectors, it could offer advantages for ULDM searches in frequency regions where detectors are limited. Specifically, in the bands of LISA, where the coherent time scale is relatively long as $\tau_{\rm coh}\sim 290\,{\rm yr}$, correlations between data at different times are expected to be effective.

\begin{acknowledgements}
    We thank Takahiro Tanaka, Hidetoshi Omiya, Soichiro Morisaki, and LIGO-Virgo-KAGRA Collaboration members for their helpful discussion. YM and SM are grateful for the hospitality of Perimeter Institute where part of this work was carried out. Research at Perimeter Institute is supported in part by the Government of Canada through the Department of Innovation, Science and Economic Development and by the Province of Ontario through the Ministry of Colleges and Universities. KA would like to thank the Institute of Cosmos Sciences of the University of Barcelona for their hospitality during his visit. This work is supported by the establishment of university fellowships towards the creation of science technology innovation (Y.M.), Japan Society for the Promotion of Science (JSPS) Overseas Challenge Program for Young Researchers (Y.M.), JSPS Grants-in-Aid for Scientific Research No.~21J01383 (H.T.), No.~22K14037 (H.T.), No.~20K14468 (K.A.), No.~18K13537 (T.F.), No.~20H05854 (T.F.), and by World Premier International Research Center Initiative, MEXT, Japan.
\end{acknowledgements}  

\appendix

\section{ORF for the finite-time traveling effect}
\label{sec:tildeORF}

The ORFs for finite-time traveling effect for $X=\{\mathsf{T},\mathsf{L}\}$, which is defined in \eqref{eq:GammaTLdef}, are represented as
\begin{align}
    {\Gamma}_{IJ}^{X}=D_I^i D_J^j\tilde{\Gamma}^{X}_{ij}\,,
    \label{eq:Gamma_X_IJ}
\end{align}
where $D_{I}^i$ and $D_{J}^i$ are the detector vectors defined by \eqref{eq:detectorvector}, and $\tilde{\Gamma}^{X}_{ij}$ have been defined by
\begin{align}
    \tilde{\Gamma}^{\mathsf{T}}_{ij}&\equiv\frac{3}{4}\int\frac{d^2{\bm \Omega}}{4\pi}\left(e^x_i(\hat{\bm{\Omega}})e^x_j(\hat{\bm{\Omega}})+e^y_i(\hat{\bm{\Omega}})e^y_j(\hat{\bm{\Omega}})\right)\,,\label{eq:tildeGammaT}
    \\
    \tilde{\Gamma}^{\mathsf{L}}_{ij}&\equiv\frac{3}{2}\int\frac{d^2{\bm \Omega}}{4\pi}e^z_i(\hat{\bm{\Omega}}) e^z_j(\hat{\bm{\Omega}})\,.\label{eq:tildeGammaL}
\end{align}
Here, $\hat{\bm d}$ denotes a unit vector in the direction from the detector $J$ to $I$. Thanks to the symmetry of index as $\Gamma_{ij}^X=\Gamma_{ji}^X$, $\tilde{\Gamma}_{ij}$ can be decomposed as
\begin{align}
    \tilde{\Gamma}^{X}_{ij}&=C^X_1\delta_{ij}+C^X_2\hat{d}_i \hat{d}_j\,.
    \label{eq:GammaTildeij}
\end{align}
We define the contracted quantities of $\tilde{\Gamma}^X_{ij}$ as
\begin{align}
    q^X_1\equiv\tilde{\Gamma}^X_{ij}\delta^{ij}\,,\quad
    q^X_2\equiv\tilde{\Gamma}^X_{ij}\hat{d}^i\hat{d}^j\,.
\end{align}
Thus, the coefficients $(C^X_1,C^X_2)$ relate to $(q^X_1,q^X_2)$ by
\begin{align}
    \left(
    \begin{matrix}
        q^X_1\\q^X_2
    \end{matrix}
    \right)
    =
    \left(
    \begin{matrix}
        3 & 1\\
        1 & 1
    \end{matrix}
    \right)
    \left(
    \begin{matrix}
        C^X_1\\C^X_2
    \end{matrix}
    \right)\,.\label{eq:matrix}
\end{align}
We define the polarization basis as
\begin{align}
    \hat{\bm{m}}&=
    \left(-\sin\varphi,\cos\varphi,0\right)\,,\label{eq:m}
    \\
    \hat{\bm{n}}&=
    \left(
        -\cos\theta\cos\varphi,
        -\cos\theta\sin\varphi,
        \sin\theta
    \right)\,,\label{eq:n}\\
    \hat{\bm{\Omega}}&=
    \left(
        \sin\theta\cos\varphi,
        \sin\theta\sin\varphi,
        \cos\theta
    \right)\,,\label{eq:Omega}
\end{align}
and set $\hat{\bm{d}}=(0,0,1)$. Integrating \eqref{eq:tildeGammaT} and \eqref{eq:tildeGammaL} over the celestial sphere, $(q^X_1,q^X_2)$ can be calculated as
\begin{align}
    \left(
    \begin{matrix}
        q_1^\mathsf{T}\\
        q_2^\mathsf{T}
    \end{matrix}
    \right)
    =
    \left(
    \begin{matrix}
        q_1^\mathsf{L}\\
        q_2^\mathsf{L}
    \end{matrix}
    \right)
    =
    \left(
    \begin{matrix}
        \frac{3}{2}\\
        \frac{1}{2}
    \end{matrix} 
    \right)\,.
    \label{eq:q1q2}
\end{align}
From \eqref{eq:GammaTildeij}, \eqref{eq:matrix}, and \eqref{eq:q1q2}, we obtain
\begin{align}
    \tilde{\Gamma}^\mathsf{T}_{ij}=\tilde{\Gamma}^\mathsf{L}_{ij}=\frac{1}{2}\delta_{ij}\,.
    \label{eq:GammaTilde}
\end{align}

The detector vector $D_I^i$ is defined as the difference between two unit vectors $\hat{\bm{X}}$ and $\hat{\bm{Y}}$, which are along each arm of detectors. These unit vectors are described by the spherical coordinate spanned by the basis $\{\bm{e}_{\hat{\theta}},\bm{e}_{\hat{\phi}},\hat{\bm{d}}\}$ as
\begin{align}
    \hat{\bm X}&=\cos\left(\sigma-\frac{\pi}{4}\right){\bm e}_{\hat{\theta}}+\sin\left(\sigma-\frac{\pi}{4}\right){\bm e}_{\hat{\phi}}\,,\label{eq:Xhat}\\
    \hat{\bm Y}&=\cos\left(\sigma+\frac{\pi}{4}\right){\bm e}_{\hat{\theta}}+\sin\left(\sigma+\frac{\pi}{4}\right){\bm e}_{\hat{\phi}}\,.
    \label{eq:Yhat}
\end{align}
Substituting \eqref{eq:GammaTilde}, \eqref{eq:Xhat}, and \eqref{eq:Yhat} into \eqref{eq:Gamma_X_IJ}, we get
\begin{align}
    \Gamma^{\mathsf{T}}_{IJ}=\Gamma^{\mathsf{L}}_{IJ}=\cos^2\left(\frac{\beta}{2}\right)\cos(2\delta)
    -\sin^2\left(\frac{\beta}{2}\right)\cos(2\Delta)\,,
\end{align}
where we have employed the parameters $\{\beta,\delta,\Delta\}$, which is defined by \eqref{eq:notation_sigma}.

\section{ORF for the Cross Term}
\label{sec:app2}

For spin-0 and spin-1 ULDM signal,
the cross-correlation includes the cross terms of $h_{\rm space}$ and $h_{\rm time}$ such as $\left<h_{\rm space}^*(f)h_{\rm time}(f')\right>$. Its ORF is defined by
\begin{align}
    \calG^\mathsf{T}_{IJ}&\equiv\int\frac{d^2\hat{\bm{\Omega}}}{4\pi}(F^x_{I}(\hat{\bm{\Omega}})G^x_{J}(\hat{\bm{\Omega}})+F^{y}_{I}(\hat{\bm{\Omega}})G^{y}_{J}(\hat{\bm{\Omega}}))\,,
    \\
    \calG^\mathsf{L}_{IJ}&\equiv\int\frac{d^2\hat{\bm{\Omega}}}{4\pi}F^z_{I}(\hat{\bm{\Omega}})G^z_{J}(\hat{\bm{\Omega}})\,.
\end{align}
These ORFs can be rewritten in the form of
\begin{align}
    \calG^X_{IJ}=D_{I}^{ij}D_{J}^{k}\tilde{\calG}_{ijk}^X\,,
    \quad
    (X\in\{\mathsf{T},\mathsf{L}\})
\end{align}
with
\begin{align}
    \tilde{\calG}^\mathsf{T}_{ijk}&=\int \frac{d^2\hat{\bm{\Omega}}}{4\pi }(e^x_{ij}(\hat{\bm{\Omega}})e^x_{k}(\hat{\bm{\Omega}})+e^y_{ij}(\hat{\bm{\Omega}})e^y_{k}(\hat{\bm{\Omega}}))\,,
    \\
    \tilde{\calG}^\mathsf{L}_{ijk}&=\int \frac{d^2\hat{\bm{\Omega}}}{4\pi}e^z_{ij}(\hat{\bm{\Omega}})e^z_{k}(\hat{\bm{\Omega}})\,.
\end{align}
Thanks to the symmetry of induces as ${\calG}_{ijk}^X={\calG}_{jik}^X$, $\tilde{\calG}_{ijk}$ is represented by the combinations of $\delta_{ij}$ and $\hat{d}_i$ as
\begin{align}
    \tilde{\calG}^X_{ijk}=C^X_1\delta_{ij}\hat{d}_{k}+C^X_2\delta_{k(i}\hat{d}_{j)}+C^X_3\hat{d}_i \hat{d}_j \hat{d}_k\,.\label{eq:GtildeXijk}
\end{align}
We define $(q_1^X,q_2^X,q_3^X)$ by
\begin{align}
    q^X_1&\equiv\tilde{\calG}^X_{ijk}\delta^{ij}\hat{d}^k\,,\label{eq:q1}
    \\
    q^X_2&\equiv\tilde{\calG}^X_{ijk}\delta^{ik}\hat{d}^j\,,\label{eq:q2}
    \\
    q^X_3&\equiv\tilde{\calG}^X_{ijk}\hat{d}^i\hat{d}^j\hat{d}^k\,.\label{eq:q3}
\end{align}
From \eqref{eq:GtildeXijk}, $(q_1^X,q_2^X,q_3^X)$ relate to $(C_1^X,C_2^X,C_3^X)$ as
\begin{align}
    \left(
    \begin{matrix}
        C_1^X\\C_2^X\\C_3^X
    \end{matrix}
    \right)
    =
    \frac{1}{2}\left(
    \begin{matrix}
        1 & 0 & -1\\
        0 & 2 & -2\\
        -1 & -2 & 5
    \end{matrix}
    \right)
    \left(
    \begin{matrix}
        q_1^X\\q_2^X\\q_3^X
    \end{matrix}
    \right)\,.
\end{align}
Applying the polarization basis defined by \eqref{eq:m}, \eqref{eq:n}, and \eqref{eq:Omega}, the angular integral in \eqref{eq:q1}, \eqref{eq:q2}, and \eqref{eq:q3} can be performed. As a result, we find $(q^X_1,q^X_2,q^X_3)=(0,0,0)$ for $X=\{\mathsf{T},\mathsf{L}\}$. Thus, the ORFs for the cross term vanish as
\begin{align}
    \calG^\mathsf{T}_{IJ}=\calG^\mathsf{L}_{IJ}=0\,.
\end{align}

\bibliography{ref}

\end{document}